\documentclass[11pt]{article}
\usepackage{epsfig}

\begin{document}

\def\xslash#1{{\rlap{$#1$}/}}
\def \p {\partial}
\def \dd {\psi_{u\bar dg}}
\def \ddp {\psi_{u\bar dgg}}
\def \pq {\psi_{u\bar d\bar uu}}
\def \jpsi {J/\psi}
\def \psip {\psi^\prime}
\def \to {\rightarrow}
\def\bfsig{\mbox{\boldmath$\sigma$}}
\def\DT{\mbox{\boldmath$\Delta_T $}}
\def\xit{\mbox{\boldmath$\xi_\perp $}}
\def \jpsi {J/\psi}
\def\bfej{\mbox{\boldmath$\varepsilon$}}
\def \t {\tilde}
\def\epn {\varepsilon}
\def \up {\uparrow}
\def \dn {\downarrow}
\def \da {\dagger}
\def \pn3 {\phi_{u\bar d g}}

\def \p4n {\phi_{u\bar d gg}}

\def \bx {\bar x}
\def \by {\bar y}

\begin{center}
{\Large\bf   QCD Corrections of All Structure Functions in Transverse Momentum Dependent Factorization for Drell-Yan Processes }
\par\vskip20pt
J.P. Ma$^{1,2}$ and G.P. Zhang$^{2}$     \\
{\small {\it
$^1$ Institute of Theoretical Physics, Academia Sinica,
P.O. Box 2735,
Beijing 100190, China\\
$^2$ Center for High-Energy Physics, Peking University, Beijing 100871, China
}} \\
\end{center}
\vskip 1cm
\begin{abstract}
We study the one-loop correction in Transverse-Momentum-Dependent(TMD) factorization for Drell-Yan processes at small transverse momentum of the lepton pair. We adopt the so-called subtractive approach,  in which one can systematically construct contributions for subtracting long-distance effects represented by diagrams. The perturbative parts are obtained after the subtraction.
We find that the perturbative coefficients of all structure functions in TMD factorization at leading twist are the same.
The perturbative parts can also be studied with scattering of partons instead of hadrons.
In this way, the factorization of  many structure functions can only be examined by studying  the scattering of multi-parton states, where there are many diagrams. These diagrams have no similarities to those treated in the subtractive approach.
As an example, we use existing results of one structure function responsible for Single-Spin-Asymmetry, to show
that these diagrams in the scattering of multi-parton states are equivalent to those treated in the subtractive approach
after using Ward identity.

\vskip 5mm
\noindent
\end{abstract}

\par
\noindent
{\large\bf 1. Introduction}
\par\vskip5pt
QCD factorization is an important concept  for studying high energy scattering, in which
both long-distance- and short-distance effects exist. With a proven factorization one can
consistently separate short-distance effects from long-distance effects.
The separated short-distance effects can be safely calculated with perturbative QCD.
The long-distance effects can be represented by matrix elements, which are consistently
defined with QCD operators\cite{Fac}.  This fact allows us not only to make predictions, but also to explore the inner structure of hadrons through determining these matrix elements from experiment.

\par
There are two types of QCD factorizations for inclusive processes. One is of collinear factorization,
in which one neglects the transverse motion of partons inside hadrons at the leading twist or at the leading power.  Another one is of Transverse-Momentum-Dependent(TMD) factorization, where
one takes the transverse momenta of partons into account at the leading power.  Using this factorization allows one to study the transverse motion of partons in hadrons and hence
to obtain 3-dimensional information about the inner structure of hadrons.
TMD factorization is applicable for processes in certain kinematical regions.  E.g., in Drell-Yan processes, the TMD factorization can only be used if the lepton pair has a small transverse
momentum $q_\perp$ which is much less than the invariant mass $Q$ of the lepton
pair.  In this work we focus on one-loop correction in TMD factorization for Drell-Yan processes.
\par
TMD factorization has been first studied in the case where a nearly back-to-back hadron pair
is produced in $e^+e^-$-annihilations\cite{CS}. A factorization theorem in this case is established.
Later, such a factorization has been established or examined in Drell-Yan processes\cite{CSS},
Semi-Inclusive Deeply Inelastic Scattering(SIDIS)\cite{JMY,CAM}, and has been extended to the polarized case\cite{JMYP}. The established TMD factorization in SIDIS and Drell-Yan processes
only involves TMD quark distributions at the order of leading power $q_\perp/Q$.
There exist TMD gluon distributions. These distributions can be extracted from inclusive processes in hadron collision like 
Higgs-production\cite{JMYG,SecG1}, quarkonium production \cite{BoPi,MWS} and two-photon production\cite{QSW}.   It should be noted that studies of TMD factorization
will not only help to explore the inner structure of hadrons, but also provide a framework
for resummtion of large log terms with $\ln q_\perp/Q$ in perturbative expansion with $q_\perp \gg
\Lambda_{QCD}$. The classical example is for Drell-Yan processes studied in \cite{CSS}.
\par
Unlike parton distributions in collinear factorization at leading twist, there are many TMD parton
distributions at leading twist. Structure functions, e.g., in Drell-Yan processes, are factorized with these distributions.
The perturbative coefficients at tree level in TMD factorization can be  easily derived.
However, for reliable predictions one needs to know higher-oder corrections in the factorization. This
is also important for giving physical predictions of experiments performed at different energy scales,
since the dependence on the scales of perturbative coefficients appears beyond tree-level.
\par
In Drell-Yan processes,  one-loop correction of some structure functions can be obtained
by studying partonic scattering and TMD parton distributions of a single parton, where one replaces each hadron with a single parton,
i.e., the scattering $a+b \to \gamma^* +X$ with $a$ or $b$ as a single parton state.
The one-loop corrections of the studied
structure functions in \cite{JMYP} are in fact obtained in this way.  But this approach
for obtaining higher-order corrections does not work for many other structure functions,
e.g., the structure function for Single transverse-Spin Asymmetry(SSA) in the case that
one initial hadron is transversely polarized. This structure function
is factorized with the TMD parton distribution, called Sivers function\cite{Sivers}.
If one replaces the transversely polarized hadron with a transversely polarized quark,
one will always have zero  results  for the structure function and the Sivers function, because the chirality
of a massless quark is conserved in perturbative QCD. Therefore, one needs to use multi-parton state
instead of a single parton state to study those structure functions.
Such a study for SSA has been done mainly in the framework of collinear factorization in \cite{MS1,MS2,MS22}.  The approach with multi-parton states has provided a different way
to solve some discrepancies in collinear factorization of SSA\cite{MS22,TW3EVO}.
\par

In principle one can use these multi-parton states to study higher-order
corrections in TMD factorization. Since scattering  with  multi-parton states
is more complicated, the one-loop correction is difficult to obtain, because too many diagrams
are involved.
In this work we use the so-called subtractive approach to study the problem. The approach
is based on diagram expansion and explained in \cite{JC1,JC2,JC3}.
In general, it is relative easy to find the leading order contribution to structure functions in the factorized form.
At the next-to-leading or one-loop order, the diagrams,
which give possible contributions, contain in general
contributions from TMD parton distributions, which are of long-distance effects and need to be subtracted for obtaining
the one-loop perturbative coefficients. In the subtractive approach
one can systematically construct such subtractions in terms of diagrams.  A comparison
of the two approaches can be noticed in the following: In the approach with multi-parton states one explicitly  calculates in detail the contributions
of structure functions and the correspond contributions of TMD parton distributions  for the subtraction.  In the subtractive
approach one only calculates in detail the contributions to the structure functions subtracted
with the contributions of TMD parton distributions. At leading twist of TMD factorization, the symmetric
part of the hadronic tensor has 24 structure functions\cite{TaMu,AMS}.
With the work presented here, it turns out that the one-loop correction is the same
for all structure functions.
This result can be generalized beyond one-loop order.
\par
It may be difficult to understand why the one-loop correction is the same for all
structure functions. Taking SSA factorized with Sivers function as an example,  the diagrams treated in the subtractive
approach have no similarity to those diagrams treated with multi-parton states.  We will make a comparison
for a part of existing results for SSA to show that the contribution of the studied part is the same
obtained from diagrams in the subtractive approach.
This is in fact a consequence of Ward identity.
\par
In Drell-Yan processes, the interpretation
of the small $q_\perp$ is that it is partly generated with the transverse momentum of incoming
partons from hadrons. In TMD factorization, as we will see in the subtractive approach,
one momentum-component of partons is set to be zero as an approximation. This may
give the impression that one deals here with scattering of off-shell partons and hence
brings up the question if the TMD factorization is gauge-invariant. We will discuss this
problem and show that the factorization is gauge-invariant.
\par
Our work is organized as in the following: In Sect. 2. we give our notation and derive
the tree-level result. In Sect. 3. we discuss  the issue of gauge invariance mentioned in the above.
In Sect. 4. and Sect. 5. we analyse the one-loop contributions in the factorization with the subtractive
approach and give our main results. In Sect. 6. we make a comparison with a part of results
derived with the subtractive approach and the existing results calculated with multi-parton states.
Sect.7. is our summary.   Detailed results for all factorized 24 structure functions are given in the Appendix.

\par
\par\vskip10pt

\noindent
{\large \bf 2. Notations and Tree-Level Results}
\par\vskip5pt

We consider the Drell-Yan process:
\begin{equation}
  h_A ( P_A) + h_B(P_B ) \to  \gamma^* (q) + X \to \ell^-  + \ell ^+ + X.
\end{equation}
We will use the  light-cone coordinate system, in which a
vector $a^\mu$ is expressed as $a^\mu = (a^+, a^-, \vec a_\perp) =
((a^0+a^3)/\sqrt{2}, (a^0-a^3)/\sqrt{2}, a^1, a^2)$ and $a_\perp^2
=(a^1)^2+(a^2)^2$.
We take a light-cone coordinate system in which:
\begin{equation}
P_{A}^\mu \approx  (P_A^+, 0, 0,0),  \ \ \ \  P_B ^\mu \approx ( 0, P_B^-, 0,0).
\end{equation}
$h_A$ moves in the $z$-direction, i.e., $P_A^+$ is the large component.
The lepton pair or the virtual photon carries the momentum $q$ with $q^2 =Q^2$.
We will study the case with $q^2_\perp \ll Q^2$ and that the two initial hadrons are of spin-1/2. The two hadrons
are polarized. The polarization of hadron $A$ can be described by the helicity $\lambda_{A}$
and a transverse spin vector $S_{A}^\mu=(0,0,S_A^1, S_A^2)$.  The polarization of hadron $B$
is described by $\lambda_B$ and $S_{B}^\mu$. For convenience
we also introduce two light-cone vectors: $n^\mu =(0,1,0,0)$ and $l^\mu = (1,0,0,0)$, and
two transverse tensors:
\begin{equation}
  g_\perp^{\mu\nu} = g^{\mu\nu} - n^\mu l^\nu - n^\nu l^\mu,
  \ \ \ \ \ \
  \epsilon_\perp^{\mu\nu} =\epsilon^{\alpha\beta\mu\nu}l_\alpha n_\beta,
\end{equation}
The relevant hadronic tensor  for Drell-Yan processes is defined as:
\begin{eqnarray}
W^{\mu\nu}  = \sum_X \int \frac {d^4 x}{(2\pi)^4}  e^{iq \cdot x} \langle h_A (P_A), h_B(P_B)  \vert
    \bar q(0) \gamma^\nu q(0) \vert X\rangle \langle X \vert \bar q(x) \gamma^\mu q(x) \vert
     h_B(P_B),h_A (P_A)  \rangle.
\label{HW}
\end{eqnarray}
The tensor can be decomposed into various structure functions.  In this work we will only give
results for those structure functions which receive leading-twist contributions in TMD factorization.

\par
Taking hadrons as bound states of partons, i.e., quarks and gluons,  the
scattering of hadrons, hence the hadronic tensor can be represented by Feynman diagrams.
Regardless how these diagrams are complicated, one can always divide
each diagram into three parts: One part contains the insertion of two electromagnetic currents as
indicated in Eq.(\ref{HW}). The other two parts are related to the hadron $h_A$ or $h_B$.
The three parts are connected with parton lines.  An example is given in Fig.1a. In Fig.1a,
the middle part contains the two electromagnetic vertices, the lower part
is associated with $h_A$  and the upper part is associated with $h_B$.
Two quark lines connect the middle part with  the part of $h_A$ and
two antiquark lines connect the middle part with the part of $h_B$.
The parton lines from the part of $h_A$ only denote the contraction of Dirac- and color
indices with the middle part, and the momentum flow into the middle part.  The propagators  associated with the parton lines are in the part
of $h_A$. The same is also for the part of $h_B$.  The middle part can be classified with the order of $\alpha_s$. E.g., at tree-level the middle part of Fig.1a is given by Fig.1b.
Hereafter, we denote the part of $h_A$ or $h_B$ as sum of all possible diagrams for a given
middle part. In this work we use Feynman gauge.

\begin{figure}[hbt]
\begin{center}
\includegraphics[width=14cm]{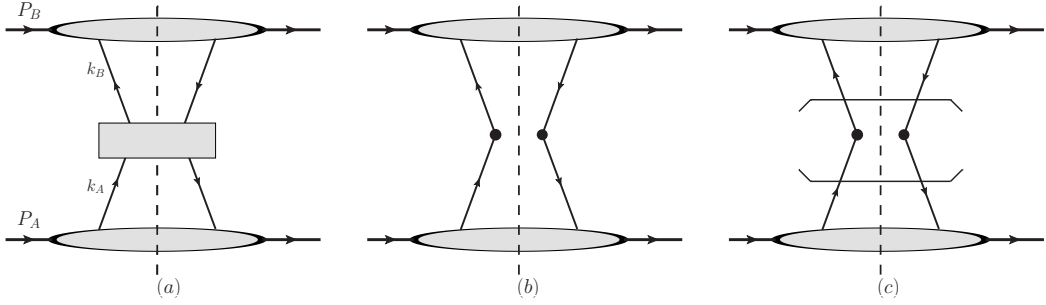}
\end{center}
\caption{(a): The general structure of diagrams. (b): The leading order diagram.
 (c): The hooked lines denote the approximation applied for Fig.1b. The black dots denote the insertion of the electric current operators.   }
\label{Feynman-dg1}
\end{figure}


\par
The part of $h_A$ and of $h_B$  in Fig.1a can be identified as:
\begin{eqnarray}
\Gamma_{ji}(P_A,k_A ) &=& \int \frac{ d^4\xi} {(2\pi)^4} e^{-i \xi \cdot k_A}
\langle h_A (P_A) \vert  \left [ \bar q(\xi) \right ]_i
 \left [ q (0)\right ]_j \vert h_A (P_A) \rangle ,
\nonumber\\
 \bar \Gamma_{ij}(P_B,k_B) &=& \int \frac{ d^4\xi} {(2\pi)^4} e^{-i \xi \cdot k_B}
\langle h_B (P_B) \vert  \left [  q(\xi) \right ]_i
 \left [ \bar q (0)\right ]_j \vert h_B (P_B) \rangle ,
\end{eqnarray}
where $ij$ stand for Dirac- and color indices. We denote the middle part in Fig.1a as
$H^{\mu\nu}_{ij,kl} (k_A,k_B,q)$, the contribution of Fig.1a as:
\begin{equation}
W^{\mu\nu} (P_A,P_B,q) \biggr\vert_{1a} = \int d^4 k_A d^4 k_B  \Gamma _{ji} (P_A,k_A)
H^{\mu\nu}_{ij, lk} (k_A,k_B,q)  \bar  \Gamma_{kl}(P_B,k_B).
\end{equation}

\par
To factorize the contribution from Fig.1a with $q_\perp \sim \Lambda_{QCD}$ and $q_\perp /Q \ll 1$, especially at tree level, certain approximations
can be made. Because we are interested in the kinematical region of $q_\perp \sim \Lambda_{QCD}$, the momenta of $k_{A\perp}^\mu$ and
$k_{B\perp}^\mu$ can not be arbitrarily large. They are restricted as $k^{\mu}_{A\perp}\sim k_{B\perp}^\mu\sim \Lambda_{QCD}$.
Here a detailed discussion is needed to clarify what is in fact included in the hadronic parts in Fig.1b. This is also important
for the comparison with the detailed calculation in Sect. 6. For this we take $\Gamma(P_A,k_A)$ as an example. In principle there can be the
case that $\Gamma$
receives contributions from large $k_{A\perp}^\mu\sim Q $ and also large $k_A^{-}\sim Q$ in Fig.1a. But these contributions can be calculated with perturbation theory. They need to be factorized out from $\Gamma$ and will in general give power suppressed contributions to $W^{\mu\nu}$
beyond tree-level.
Therefore, the dominant contributions only come from the case that $\Gamma(P_A, k_A)$ is only characterized with the energy scale $\Lambda_{QCD}$, i.e.,
$k_A^2\sim P_A^2 \sim \Lambda_{QCD}^2$. Hence one has $k_{A\perp}^\mu\sim \Lambda_{QCD}$ and $k_{A}^-\sim \Lambda^2_{QCD}$.
In other word, $\Gamma(P_A,k_A)$ in Fig.1b is the sum of all diagrams with $k_A^\mu \sim (1,\lambda^2,\lambda,\lambda)$ with
$\lambda =q_\perp/Q$. Similarly, one can also find that for $\bar \Gamma (P_B,k_B)$ in Fig.1b one has $k_B^\mu \sim (\lambda^2,1,\lambda,\lambda)$.
\par
With the above discussion one can find the space-time picture of the hadronic matrix element in $\Gamma$.
The $\xi^+$-dependence is characterized by the small scale $\Lambda^2_{QCD}$.  The  $\xi^-$-dependence is characterized by the scale $P_A^+ \gg \Lambda_{QCD}$, and the $\xi_\perp$-dependence is characterized by the scale
$\Lambda_{QCD}$. Therefore, we can first neglect the $\xi^+$-dependence. This is equivalent
to take the leading result by expanding $H(k_A,k_B,q)$ in $k_A^-$.  In $\bar\Gamma$ we neglect the $\xi^-$-dependence of the
hadronic matrix element.
Another approximation can be made is that the leading contributions are only given
by the matrix elements containing the good component of quark fields.
One can always decompose a quark field as:
\begin{eqnarray}
   q(x) = \frac{1}{2} \gamma^+ \gamma^- q(x) + \frac{1}{2} \gamma^- \gamma^+ q(x).
\end{eqnarray}
For $\Gamma $ the first term is the good component, the second term can be solved with equation of motion and gives a power-suppressed contribution. For $\bar \Gamma $ the second term is the
good component. After making  these approximations,  we can write the two parts
as:
\begin{eqnarray}
\Gamma_{ij}(P_A,k_A ) &\approx&  \delta (k_A^-) {\mathcal M}_{ij}(x, k_{A\perp} ) +\cdots , \quad\quad
 \bar \Gamma_{ij}(P_B,k_B) \approx  \delta (k_B^+)  \bar  {\mathcal M}_{ij}(x, k_{B\perp} ) + \cdots,
\nonumber\\
{\mathcal M}_{ij}(x, k_{A\perp})   &=&  \int \frac{ d\xi^-d^2\xi_\perp } {(2\pi)^3} e^{-i \xi \cdot \tilde k_A}
\langle h_A (P_A) \vert  \left [ \bar q(\xi) \right ]_j
 \left [ q (0)\right ]_i \vert h_A (P_A) \rangle \biggr\vert_{\xi^+ =0} ,
 \nonumber\\
 \bar {\mathcal M}_{ij}(x, k_{B\perp} ) &=&   \int \frac{ d\xi^+ d^2\xi_\perp } {(2\pi)^3} e^{-i \xi \cdot \tilde k_B}
\langle h_B (P_B) \vert  \left [  q(\xi) \right ]_i
 \left [ \bar q (0)\right ]_j \vert h_B (P_B) \rangle \biggr\vert_{\xi^-=0} ,
\label{DENM}
\end{eqnarray}
with
\begin{equation}
\tilde k_A^\mu = (x P_A^+, 0, k_A^1, k_A^2) , \quad\quad \tilde k_B^\mu =(0,x P_B^-, k_B^1,k_B^2).
\end{equation}
In Eq.(\ref{DENM}), the $\cdots$ stand for higher-twist- or power-suppressed contributions. ${\mathcal M}$
and ${ \bar {\mathcal  M}}$ are of leading-twist- or leading power. The quark fields in ${\mathcal M}$
or ${\bar {\mathcal M}}$ are correspondingly good components.
Therefore, we always have:
\begin{equation}
\gamma^+  \bar {\mathcal M}  =  \bar {\mathcal M} \gamma^+ =0, \quad\quad \gamma^- {\mathcal M} =  {\mathcal M} \gamma^- =0.
\label{MLC}
\end{equation}
This property will help us to extract the contributions of TMD parton distributions as we will see later.
The approximation made in the above is valid for the case that the transverse momentum $q_\perp$ of the lepton
pair is at order $\Lambda_{QCD}$.
The correction of the approximation to the hadronic tensor is at the order $q_\perp/Q$ or $q_\perp^2/Q^2$ relative to the
leading order.
For $q_\perp \gg \Lambda_{QCD}$ one can make a further approximation by neglecting
or expanding the $\xi_\perp$-dependence in hadron matrix elements
in $\Gamma$ or $\bar\Gamma$.  This will lead to collinear factorizations.
\par

\par
\par
We first consider the leading order given by Fig.1b. The middle part can be then given explicitly:
\begin{equation}
  H^{\mu\nu}_{ij,lk} (k_A,k_B,q) =   \delta^4 ( k_A+k_B -q)  \biggr [ \gamma^\mu \biggr ]_{lj}
    \biggr [ \gamma^\nu \biggr ]_{ik}.
\end{equation}
Using the above approximated results for $\Gamma$ and $\bar\Gamma$,  one obtains
the hadronic tensor at leading order of $\alpha_s$ as
\begin{eqnarray}
W^{\mu\nu}    \approx   \int d^2 k_{A\perp} d^2 k_{B_\perp}{\rm Tr} \biggr  [ \gamma^\mu  {\mathcal M}(x, k_{A\perp} )   \gamma^\nu \bar{\mathcal M}(y, k_{B\perp} )  \biggr ]
  \delta^2 ( k_{A\perp}+k_{B\perp} -q_\perp ) ,
\label{LTREE}
\end{eqnarray}
with $q^+ = x P_A^+$ and $q^- = y P_B^-$.
According to the notation in \cite{JC1}, we denote the approximations made for Fig.1b
to derive the above result represented by Fig.1c, it is just Fig.1b with the hooked lines. The hooked line
in the lower part denotes the approximations made for $\Gamma$ and the hooked line
in the upper part denotes the approximations made for $ \bar \Gamma$, as indicated in Eq.(\ref{DENM}).
In the above we have worked out the contribution with a quark from $h_A$ and an antiquark from $h_B$. Similarly, one can work out the case where a quark is from $h_B$ and an antiquark is from $h_A$.
\par
The density matrix ${\mathcal M}$ can be decomposed into various TMD parton distributions.
The decomposition has been studied in \cite{TMDMT,TMDBM,TMDGMS,BDGM}. At leading twist, the decomposition is: 
\begin{eqnarray}
{\mathcal M}_{ij}(x,k_\perp )  &=& \frac{1}{2N_c}\left[ f_1(x,k_\perp)\gamma^- - f_{1T}^\perp(x,k_\perp)\gamma^-\epsilon_\perp^{\mu\nu} k_{\perp\mu}S_{A\nu}\frac{1}{M_A}\right.\nonumber\\
&&+g_{1L}(x,k_\perp)\lambda_A\gamma_5\gamma^- -g_{1T}(x,k_\perp)\gamma_5\gamma^- k_\perp\cdot S_{A}\frac{1}{M_A}\nonumber\\
&&+h_{1T}(x,k_\perp)i\sigma^{-\mu}\gamma_5S_{A\mu}+h_{1}^\perp(x,k_\perp)\sigma^{\mu-}k_{\perp\mu}\frac{1}{M_A}
+h_{1L}^\perp(x,k_\perp)\lambda_A i\sigma^{-\mu}\gamma_5 k_{\perp\mu}\frac{1}{M_A}\nonumber\\
&&\left. -h_{1T}^\perp(x,k_\perp)i\sigma^{-\mu}\gamma_5 k_{\perp\mu}k_\perp\cdot S_A\frac{1}{M_A^2}\right]_{ij}.
\label{DECM}
\end{eqnarray}
There are 8 TMD parton distributions at leading twist. $M_A$ is the mass of $h_A$.
In the above
$f_{1T}^\perp$ and $h_1^\perp$ are odd under naive time-reversal transformation. $f_{1T}^\perp$ is the Sivers function,
$h_1^\perp$ is called as Boer-Mulders function. For the decomposition we have implicitly assumed that the gauge links discussed
in the next section is added in ${\mathcal M}$.
\par
Similarly one has the decomposition for ${\bar{\mathcal M}}$ as:
\begin{eqnarray}
\bar {\mathcal M}_{ij}(x,k_\perp )  &=& \frac{1}{2N_c}\left[ \bar{f}_1(x,k_\perp)\gamma^+ + \bar{f}_{1T}^\perp(x,k_\perp)\gamma^+\epsilon_\perp^{\mu\nu} k_{\perp\mu}S_{B\nu}\frac{1}{M_B}\right.
\nonumber\\
&&-\bar{g}_{1L}(x,k_\perp)\lambda_B\gamma_5\gamma^+ +\bar{g}_{1T}(x,k_\perp)\gamma_5\gamma^+ k_\perp\cdot S_{B}\frac{1}{M_B}
\nonumber\\
&&+\bar{h}_{1T}(x,k_\perp)i\sigma^{+\mu}\gamma_5S_{B\mu}+\bar{h}_{1}^\perp(x,k_\perp)\sigma^{\mu+}k_{\perp\mu}\frac{1}{M_B}
+\bar{h}_{1L}^\perp(x,k_\perp)\lambda_B i\sigma^{+\mu}\gamma_5 k_{\perp\mu}\frac{1}{M_B}
\nonumber\\
&&\left. -\bar{h}_{1T}^\perp(x,k_\perp)i\sigma^{+\mu}\gamma_5 k_{\perp\mu}k_\perp\cdot S_B\frac{1}{M_B^2}\right]_{ij}.
\label{DECMB}
\end{eqnarray}
It should be noted that ${\mathcal M}$ and ${\bar{\mathcal M}}$ are diagonal in colour space. With
these decomposition one can work out the hadronic tensor at leading order. The results for the tensor
can be represented with structure functions and each structure function is factorized with
corresponding TMD parton distributions.

\par\vskip20pt
\noindent

\begin{figure}[hbt]
\begin{center}
\includegraphics[width=11cm]{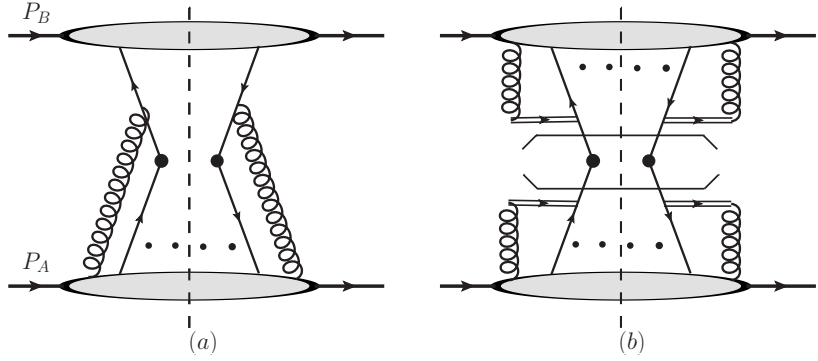}
\end{center}
\caption{(a):  Exchange of gluons between the part with $h_A$ and the antiquark-line. The number of the exchanged gluons is arbitrary denoted by the black small dots. (b): Factorization of longitudinal polarized gluons. The double lines are for gauge links.}
\label{dg2}
\end{figure}

\noindent
{\large \bf  3.  Gauge Invariance and Gauge Links}
\par\vskip5pt
In the last section we have worked out the tree-level TMD factorization by considering
the diagram Fig.1b. In this diagram or Fig.1a, there are only two parton lines connecting
the middle part with the part of $h_A$, and two parton lines connecting the middle part
with the part of $h_B$. From argument of power-counting, if the connection is made by more
parton lines in Fig.1a or 1b, the contributions are power-suppressed but with exceptions.
The exceptions are well-known.  If there are gluon lines connecting the middle part with
the part of $h_A$, and those lines are for $G^+$-gluons collinear to $h_A$, the resulted contributions are not power-suppressed. 
The tree-level diagram with many collinear gluons is illustrated in Fig.2a.
\par
It is well-known how those diagrams with many collinear gluons can be summed. The summation
is achieved by introducing gauge links. In this work we follow \cite{JMY} by using the gauge link
along the direction $u^\mu =(u^+,u^-,0,0)$ with $u^-\gg u^+$:
\begin{equation}
  {\mathcal L}_u(\xi, -\infty) = P \exp \left  (-i g_s \int_{-\infty}^0 d\lambda u\cdot G (\lambda u +\xi) \right ) .
\end{equation}
The diagrams like that given in Fig.2a can be summed by inserting in ${\mathcal M}$
the product  ${\mathcal L}_u (\xi, -\infty) {\mathcal L}_u^\dagger (0, -\infty) $. Similarly,
there can be diagrams with many collinear $G^-$-gluons emitted by the part of $h_B$. 
These diagrams can also be summed
by introducing the gauge link ${\mathcal L}_v$ along the direction $v^\mu =(v^+,v^-,0,0)$
with $v^+\gg v^-$. Finally, the summation can be represented by Fig.2b.
The summation is made by  re-defining:
\begin{eqnarray}
{\mathcal M}(x,k_\perp)  &=&  \int \frac{ d\xi^-d^2\xi_\perp } {(2\pi)^3} e^{ -i x \xi^- P_A^+ - i \xi_\perp \cdot k_\perp}
\langle h_A (P_A) \vert   \bar q(\xi) {\mathcal L}_u (\xi, -\infty) {\mathcal L}_u^\dagger (0, -\infty)
  q (0) \vert h_A (P_A) \rangle \biggr\vert_{\xi^+ =0} ,
 \nonumber\\
 \bar {\mathcal M} (x, k_\perp) &=&   - \int \frac{ d\xi^- d^2\xi_\perp } {(2\pi)^3} e^{-ix \xi^+ P_B^- - i  \xi_\perp \cdot k_\perp }
\langle h_B (P_B) \vert    \bar q(0) {\mathcal L}_v (0, -\infty) {\mathcal L}_v^\dagger (\xi, -\infty)  q (\xi) \vert h_B (P_B) \rangle \biggr\vert_{\xi^-=0} ,
\nonumber\\
\label{DENMG}
\end{eqnarray}
these matrices are diagonal in color-space and $4\times 4$-matrices in Dirac space, and $x\ge 0$.  With these gauge links, TMD parton distributions will not only depend on $x,k_\perp$ and the renormalization scale $\mu$ but also depend
on those parameters:
\begin{equation}
\zeta_u^2 =\frac{(2P_A \cdot u)^2}{u^2},\quad\quad\quad \zeta_v^2 =\frac{ (2 P_B\cdot v)^2}{v^2}.
\end{equation}
The dependence on these parameters is controlled by Collins-Soper equation\cite{CS}.
The Collins-Soper equations of the introduced TMD parton distributions can be found in \cite{CS,AJMY}. 
In general one needs to add in Eq.(\ref{DENMG}) gauge links along transverse direction at infinite space-time to make 
density matrices gauge invariant as shown in \cite{TMDJi}. In this work, we will take a non-singular gauge, i.e., Feynman gauge. 
In a non-singular gauge gauge links at infinite space-time vanish.    
\par
With the added gauge links the TMD parton distributions are gauge invariant. But there seems
another problem of gauge invariance related to the tree-level result in Eq.(\ref{LTREE}).
The result at the leading order seems that one can interpret the process as: One quark with
the momentum $\tilde k_A^\mu =(k_A^+, 0, k^1_{A\perp}, k^2_{A\perp})$ from $h_A$ annihilates
an antiquark with the momentum $\tilde k_B^\mu =(0, k_B^-, k^1_{B\perp}, k^2_{B\perp})$
from $h_B$ into the virtual photon. From the momenta one can realize that the quark and the antiquark
are off-shell because of $\tilde k_A^2 \neq 0$ and $\tilde k_B^2\neq 0$. One may conclude
that the result is not gauge invariant because the perturbative coefficients are extracted
from scattering amplitudes of off-shell partons.  However it can be shown as in the following that this is not the case.
\par
\begin{figure}[hbt]
\begin{center}
\includegraphics[width=10cm]{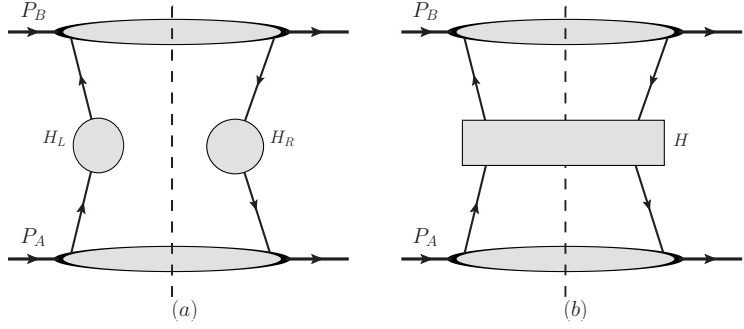}
\end{center}
\caption{(a):  A class of diagrams where no parton goes through the cut in the middle part. (b):  A class of diagrams  there are patrons crossing the cut.  }
\label{dg3}
\end{figure}
\par
\par
We introduce two momenta which are:
\begin{equation}
 \bar k_A^\mu =(k_A^+, 0, 0,0) , \quad\quad   \bar  k_B^\mu =(0, k_B^-, 0,0).
 \label{COMEN}
\end{equation}
These are momenta of on-shell partons.
Now using the properties in Eq.(\ref{MLC}) we can derive:
\begin{eqnarray}
{\rm Tr } \biggr [ \gamma^\mu {\mathcal M} \gamma^\nu \bar{\mathcal M} \biggr ]
  &=&  \frac{1}{16 ( k_A^+ k_B^-)^2 } \sum_{s_1,s_2,s_3,s_4}  \biggr [ \bar v(\bar k_B, s_1) \gamma^\mu u(\bar k_A,s_2)  \biggr ]
  \bar u(\bar k_A,s_2)
     \gamma^+ {\mathcal M} \gamma^+ u(\bar k_A,s_3)
\nonumber\\
   &&   \quad\quad   \cdot  \biggr [  \bar u(\bar k_A, s_3)\gamma^\nu v(\bar k_B,s_4) \biggr ] \bar v(\bar k_B, s_4) \gamma^- {\mathcal M} \gamma^- v(\bar k_B,s_1),
\end{eqnarray}
from this one can see that the perturbative coefficients are in fact extracted from
scattering amplitudes of on-shell partons, i.e., from the annihilation amplitude $q(\bar k_A) \bar q(\bar k_B) \to
\gamma^*(\bar q) $ with $\bar q^\mu =(q^+,q^-,0,0)$, indicated by the two terms in the two terms $[ \cdots ]$.  The effect of transverse momenta of partons are only taken into account in the momentum conservation, i.e., in the $\delta$-function $\delta^2 (k_{A\perp} + k_{B\perp} -q_\perp)$.
Therefore, the tree-level result in Eq.(\ref{LTREE}) are gauge invariant.
\par

\par
It is rather obscure to see if the tree-level result in Eq.(\ref{LTREE}) is gauge invariant from
momenta carried by patrons, because the amplitudes there are constant. If we go beyond
tree-level, we can see this more clearly. We consider a class of diagrams in which
there is no parton crossing the cut. This case is represented by Fig.3a.
After making the approximations indicated with Eq.(\ref{DENM}) the contribution from Fig.3a can be in general written as:
\begin{equation}
 W^{\mu\nu}  \sim  {\rm Tr } \biggr [ H_L (\tilde k_A, \tilde k_B, q) {\mathcal M} (\tilde k_A) H_R (\tilde k_A, \tilde k_B, q)  \bar {\mathcal M }(\tilde k_B) \biggr ] \delta^2 (k_{A\perp} +k_{B\perp}-q_\perp).
\end{equation}
The contribution to perturbative coefficients from Fig.3a is obtained by subtracting the corresponding
contribution of TMD parton distributions from the above contribution.
Since the TMD factorization is for the leading contribution in $q_\perp /Q$, one should
find the leading contribution from Fig.3a. In its contribution the $\delta$-function already
gives the leading contribution. Hence we need to expand $H_{L,R}$ in all transverse momenta.
It is easy to find the leading contribution:
\begin{equation}
W^{\mu\nu} \sim  {\rm Tr } \biggr [ H_L (\bar k_A, \bar k_B,  \bar q) {\mathcal M } (\tilde k_A) H_R (\bar k_A, \bar k_B, \bar q)  \bar{\mathcal M}(\tilde k_B) \biggr ] \delta^2 (k_{A\perp} +k_{B\perp}-q_\perp) +\cdots
\end{equation}
where $\cdots$ stand for higher order contributions in $q_\perp/Q$.
It is clear now in $ H_{L,R}  (\bar k_A, \bar k_B,  \bar q)$ the momenta of incoming partons
are  of on-shell.  Therefore, the contribution from diagrams like that in Fig.3a is gauge invariant.
This also tells us
that for leading power contribution one only needs to calculate
$ H_{L,R}  (\bar k_A, \bar k_B,  \bar q)$  with on-shell momenta. The collinear- and infrared
singularities are then regularized with dimensional regularization.
\par
We notice here that it is important and crucial to use dimensional regularization for TMD factorization  here
for collinear- and infrared singularities. In other word, one should set $k_{A\perp}^\mu=k_{B\perp}^\mu =0$
in $H_{L,R}$ before performing integrations of loop momenta.  One may think that one can keep a nonzero
but infinite small $k_{A\perp}^2$ and $k_{B\perp}^2$ to regularize  collinear- and infrared singularities. Then these singularities will appear in $H_{L,R}$ as terms with $\ln k_{A\perp}^2$
and $\ln k_{B\perp}^2$. After subtraction of contributions from TMD parton distributions,
perturbative coefficients do not contain such terms. In fact, this is not the case. This can be seen
by calculating one-loop contribution of $H_{L,R}$ with nonzero transverse momenta of partons.
The contribution will contain terms of  $\ln k^2_{A\perp} \ln k^2_{B\perp}$. The reason for existence
of such terms is that the one-loop contribution always contain double log terms. Such terms
can never be subtracted with the contributions of TMD parton distributions, because the contributions
of TMD parton distributions at one-loop do not have such terms.
 Therefore,
with nonzero  $k_{A\perp}^2$ and $k_{B\perp}^2$  in $H_{L,R}$
TMD factorization can not be made, or the TMD factorization is gauge dependent with perturbative
coefficients depending on $k_{A\perp}^2$ and $k_{B\perp}^2$ .

\par
The above discussion is for diagrams without any parton crossing the cut. There are diagrams
with partons crossing the cut like the one given in Fig.3b. For these diagrams there is no reason
to set the transverse momenta of incoming partons to be zero as the leading  approximation.
These diagrams may have problems with gauge invariance. But, in TMD factorization, the contributions from diagrams like Fig.3b will be totally subtracted, as we will explicitly show
at one-loop, and will not contribute to perturbative coefficients.
Also, the subtracted contributions are from TMD parton distributions which are now gauge invariant,
and from a gauge-invariant soft factor which will be introduced later. Therefore, there is no problem of gauge invariance with contributions from Fig.3b.
\par
\begin{figure}[hbt]
\begin{center}
\includegraphics[width=4cm]{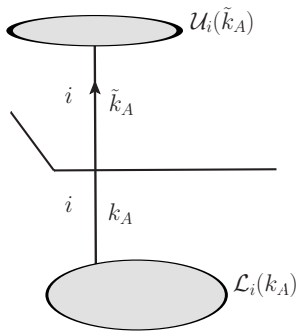}
\end{center}
\caption{The used approximation marked with a hooked line.    }
\label{dg3}
\end{figure}

\par
Before ending this section, we briefly explain the approximation denoted by hooked lines introduced
in \cite{JC1}, or the rule of theses lines here for TMD factorization.  The approximation
can be called as parton model approximation.
We consider an example with a quark, which originally comes from  $h_A$ and enters the annihilation into the
virtual photon,  as given in Fig.4.
In Fig.4 the lower part is associated with $h_A$, the upper part contains the annihilation with partons
from $h_B$. $i$ is the Dirac index.
The contribution from Fig.4 without the hooked line can be generically written as
\begin{equation}
 \Gamma =  \int d^4 k_A {\mathcal U}_i (k_A) {\mathcal L}_i(k_A) =\int d^4 k_A {\mathcal U}_i (k_A) \frac{1}{2} \left ( \gamma^- \gamma^+   + \gamma^+ \gamma^- \right )_{ij} {\mathcal L}_j(k_A),
\end{equation}
with the hooked line it means that we take the approximation for the expression as
\begin{equation}
\Gamma = \int d^2 k_{A\perp} d k_A^+ \frac{1}{2}  {\mathcal U}_i ( \tilde k_A) \left ( \gamma^- \gamma^+ \right )_{ij}   \left (  \int d k_A^- {\mathcal L}_j(k_A) \right )  + \cdots , \ \ \ \  \tilde k_A^\mu = (k_A^+, 0, k_A^1, k_A^2),
\end{equation}
where $\cdots$ stand for contributions which will give power-suppressed contributions to the hadronic tensor and are neglected.  Similarly, one can also find  the rule of hooked lines
for parton lines coming from the part of $h_B$ orginally.

\par\vskip20pt
\noindent
{\large \bf 4. One-Loop Real Correction}
\par\vskip5pt
In this section we study how the real part of one-loop correction is factorized. This part corresponds to Fig.3b. We first consider Fig.5a.
We will use this example to explain the subtractive approach in detail.
\par
\begin{figure}[hbt]
\begin{center}
\includegraphics[width=10cm]{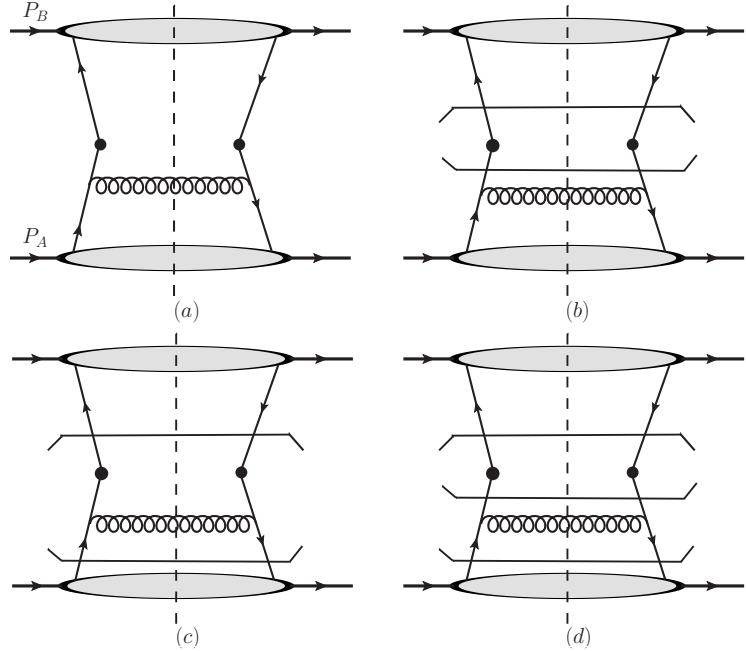}
\end{center}
\caption{(a): One of diagrams at one-loop. (b): A part of contributions from Fig.5a which is already contained in Fig.1c.
 (c): Another part of contributions from Fig.5a. (d): The part needs to be subtracted from Fig.5c.   }
\label{dg5}
\end{figure}
\par
From the topology of the diagram, we can apply the parton model approximation
in different places in the lower part of the diagram, e.g., one can apply the approximation
for the parton lines connecting to the electromagnetic vertices directly, as given in Fig.5b,
or one can also apply the approximation as given in Fig.5c and Fig.5d.
It is clearly that the contribution from Fig. 5b
is already included  in Fig.1c, i.e., in the tree-level contribution. Therefore, in order to avoid double -counting and to obtain true one-loop contribution, one needs first to subtract the contribution from Fig.5b from Fig.5a,
the remaining part will give the true one-loop correction after making the parton-model approximation for
the two parton lines from the lower bubble of $h_A$.  With the approximation Fig.5a becomes Fig.5c and Fig.5b becomes Fig.5d.
The true one-loop correction
is then given by the contribution from Fig.5c subtracted with the contribution from Fig.5d.
In the sense of the subtraction, the part of Fig.5d below the middle hooked line can be interpreted as a contribution from TMD parton distributions, because in this part all transverse momenta are at order of $\lambda$. Corresponding to the discussion for $\Gamma$ after
Eq.(6), this part is already included in $\Gamma$ or ${\mathcal M}$.

\par
We denote the momentum carried by the gluon in Fig.5 as $\tilde k_A -k$, and the gluon
is with the polarization index $\rho$.  $k$ is the momentum
entering the left electromagnetic vertex along the quark line from $h_A$.
The contribution
from Fig.5c can be easily found as:
\begin{eqnarray}
W^{\mu\nu}\biggr\vert_{5c}  &= &  - g_s^2 \int d^3 \tilde k_A d^3 \tilde k_B \frac{d^4 k}{(2\pi)^4}  2\pi \delta ((\tilde k_A -k)^2) \delta^4 (\tilde k_B +k -q)
\nonumber\\
&&  {\rm Tr} \biggr [ {\bar {\mathcal M}}(\tilde k_B) \gamma^\mu \frac{\gamma\cdot k}{ k^2 + i\varepsilon}
              \gamma^\rho T^a {\mathcal M}(\tilde k_A) \gamma_\rho T^a \frac{\gamma\cdot k}{ k^2 - i\varepsilon}  \gamma^\nu \biggr ].
\end{eqnarray}
Since we are interested in the kinematic region of $q_\perp \ll Q$, we need to find the leading contribution in the limit.
For this diagram, it is easy to find the leading contribution appears if the exchange gluon is collinear
to $P_A$, i.e.:
\begin{equation}
  (\tilde k_A -k)^\mu \sim {\mathcal O} ( 1, \lambda^2, \lambda,\lambda), \ \ \ \  \lambda^2 = \frac{q_\perp^2}{Q^2}.
\end{equation}
This also implies that $k$ is collinear  to $\tilde k_A$ because $\tilde k_A^- =0$. Using the property in Eq.(\ref{MLC})
we obtain the leading contribution in the limit $\lambda\ll 1$:
\begin{eqnarray}
W^{\mu\nu}\biggr\vert_{5c} &\approx &  - g_s^2\int d^3 \tilde k_A d^3 \tilde k_B \frac{d^4 k}{(2\pi)^4} 2\pi \delta ((\tilde k_A -k)^2) \delta^4 (\tilde k_B +k -q)
\nonumber\\
&&  {\rm Tr} \biggr [ {\bar {\mathcal M}}(\tilde k_B) \gamma^\mu_\perp \frac{\gamma_\perp \cdot k}{ k^2 + i\varepsilon}
              \gamma^\rho T^a {\mathcal M}(\tilde k_A) \gamma_\rho T^a \frac{\gamma_\perp \cdot k}{ k^2 - i\varepsilon}  \gamma^\nu_\perp  \biggr ]
               +\cdots,
\end{eqnarray}
i.e., the leading contribution has the indices $\mu,\nu$ as transverse.  The summed index $\rho$ is always transverse.
Again one can use the property in Eq.(\ref{MLC}) to re-write
the leading contribution in the form:
\begin{eqnarray}
W^{\mu\nu}\biggr\vert_{5c}      &=& \int d^2 k_{B\perp} d^2 k_\perp \delta^2 (k_{B\perp} + k_\perp - q_\perp ) {\rm Tr}\biggr \{  {\bar {\mathcal M}}(\tilde k_B) \gamma^\mu
    \biggr [ -\frac{g_s^2}{4} \int \frac{ d^3 \tilde k_A d k^-}{(2\pi)^4}  2\pi \delta ((\tilde k_A -k)^2)
 \nonumber\\
     && \cdot      \biggr (  \gamma^- \gamma^+   \frac{\gamma  \cdot k}{ k^2 + i\varepsilon}
              \gamma^\rho T^a {\mathcal M}(\tilde k_A) \gamma_\rho T^a \frac{\gamma \cdot k}{ k^2 - i\varepsilon}
               \gamma^+ \gamma^-  \biggr ) \biggr ] \gamma^\nu \biggr \} + \cdots,
\end{eqnarray}
where $\cdots$ stand for power-suppressed contributions which are neglected.
Some integrations have been done with $\delta$-functions. It gives $q^- = k_B^-$ and $q^+ =k^+$.  Now it is interesting to note that the part in $[\cdots ]$ is just the expression for the part in Fig.5d between the two hooked lines below the electromagnetic vertices. This part
can be identified as the correction of TMD parton distributions.  This fact
tells that the leading contribution from Fig.5c is the same as the contribution of Fg.5d:
\begin{equation}
 W^{\mu\nu}\biggr\vert_{5c} \approx W^{\mu\nu}\biggr\vert_{5d}.
\end{equation}
Therefore, after the subtraction Fig.5a will not contribute to perturbative coefficients in TMD factorization at one-loop.

\par
\begin{figure}[hbt]
\begin{center}
\includegraphics[width=12cm]{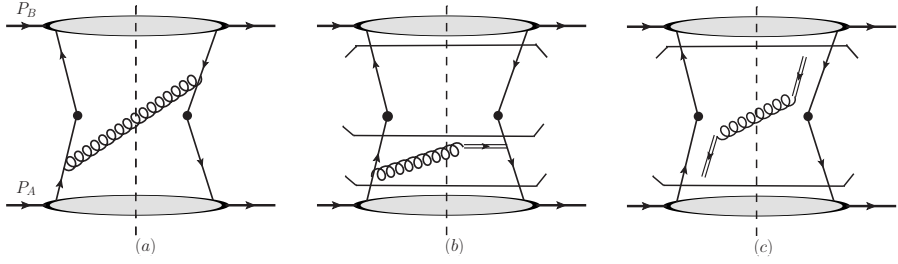}
\end{center}
\caption{ The diagrams at one-loop for the real part.    }
\label{dg6}
\end{figure}

Now we consider the contribution from Fig.6a, which also gives possible contribution at one-loop. By applying the two hooked lines in Fig.6a to parton lines nearest to the bubbles for hadrons,   we obtain  the contribution of Fig.6a as:
\begin{eqnarray}
W^{\mu\nu}\biggr\vert_{6a}  &\approx &  - g_s^2 \int d^3 \tilde k_A d^3 \tilde k_B \frac{d^4 k}{(2\pi)^4} 2\pi \delta ( (\tilde k_A -k) ^2)
   \delta^4 (\tilde k_B+ k  -q)
\nonumber\\
&&  {\rm Tr} \biggr [ {\bar {\mathcal M}}(\tilde k_B) \gamma^\mu \frac{\gamma\cdot k  }{ k^2 + i\varepsilon}
              \gamma^\rho T^a {\mathcal M}(\tilde k_A) \gamma^\nu \frac{\gamma\cdot (\tilde k_A -k -\tilde k_B  )}{ (\tilde k_B -\tilde k_A
              + k)^2 - i\varepsilon}  \gamma_\rho T^a  \biggr ],
\end{eqnarray}
here $\tilde k_A- k$ is the momentum of the exchanged gluon. It is noted that in this case the index $\rho$ can be any of $+,-,\perp$.  Again
we need to find the leading contributions from Fig.6a in $\lambda \to 0$. These leading contributions appear
in the case that $\tilde k_A -k$ is collinear to $P_A$ or to $P_B$, and $\tilde k_A - k$ is soft. We first consider the case that the exchanged gluon
is collinear to $P_A$, i.e., $k^\mu \sim {\mathcal O}(1,\lambda^2,\lambda,\lambda)$. The leading contribution
in this case is given by:
\begin{eqnarray}
W^{\mu\nu}\biggr\vert_{6aA}  &\approx &  - g_s \int d^3 \tilde k_A d^3 \tilde k_B \frac{d^4 k}{(2\pi)^4} 2\pi \delta ((\tilde k_A- k) ^2) \delta^4 (\tilde k_B+k  -q)
\nonumber\\
&&  {\rm Tr} \biggr [ {\bar {\mathcal M}}(\tilde k_B) \gamma^\mu \frac{\gamma^- k^+ }{ k^2 + i\varepsilon}
              \gamma^\rho T^a {\mathcal M}(\tilde k_A) \gamma^\nu \biggr ( \frac{  g_s n_\rho T^a  }{ n\cdot (\tilde k_A - k)  + i\varepsilon}  \biggr ) \biggr ]  + \cdots.
\end{eqnarray}
Here we use the subscript $A$ to denote the leading contribution from the momentum region
collinear to $P_A$.
It is well-known that there will be a light-cone singularity with the light-cone vector $n$.  To avoid
this we can replace
the vector $n$ with $u$ introduced before. We can use the property in Eq.(\ref{MLC}) to re-write the above result as:
\begin{eqnarray}
W^{\mu\nu}\biggr\vert_{6aA}       &=& \int d^2 k_{B\perp} d^2 k_{\perp}  \delta^2 (k_{B\perp} +k_{\perp}   -q_\perp) {\rm Tr} \biggr \{
       {\bar {\mathcal M}}(\tilde k_B) \gamma^\mu  \biggr [ -\frac {1}{4}\int  \frac{ d^3 k_A dk^-}{(2\pi)^4}  \delta ((\tilde k_A -k)^2)
\nonumber\\
&&    \biggr ( \gamma^- \gamma^+ \biggr ) \frac{\gamma\cdot k }{ k^2 + i\varepsilon}
              ( g_s \gamma^\rho ) T^a {\mathcal M}(\tilde k_A)  \biggr ( \frac{  g_s u_\rho T^a }{ u\cdot (\tilde k_A -k) + i\varepsilon}  \biggr ) \biggr (\gamma^+ \gamma^- \biggr ) \biggr ] \gamma^\nu   + \cdots,
\label{W6AA}
\end{eqnarray}
where  $q^- = k_B^-$ and $q^+ =k^+$.  With the result given in the form in the above,
one easily finds that this contribution is the same as that from Fig.6b. The contribution from Fig.6b
is already contained in Fig.1c and should be subtracted from Fig.6a. Hence, the considered
contribution in Eq.(\ref{W6AA}) is exactly subtracted and gives
no contribution to the factorization at one-loop. Similarly, one also finds that the leading contribution from Fig.6a
in the region where the exchanged gluon is collinear to $P_B$ is also exactly subtracted.
\par
The interesting part is from the region where the gluon is soft, i.e.,
$(\tilde k_A -k)^\mu \sim {\mathcal O}(\lambda,\lambda,\lambda,\lambda)$. To analyze this part
we make the substitution $(\tilde k_A -k)\to k$, i.e., now the gluon carries the momentum $k$. The leading
contribution from this region gives:
\begin{eqnarray}
W^{\mu\nu}\biggr\vert_{6aS}  &\approx & g_s^2 \int d^2 \tilde k_{A\perp} d^2 \tilde k_{B\perp}  \frac{d^4 k}{(2\pi)^4} 2\pi \delta (k ^2)
\delta^2 (k_{B\perp}+ k_{A\perp} -k_\perp  -q_\perp )
\nonumber\\
&&  {\rm Tr} \biggr [ {\bar {\mathcal M}}(\tilde k_B) \gamma^\mu
               T^a {\mathcal M}(\tilde k_A) \gamma^\nu T^a \biggr ] \frac{ n\cdot l} { (l\cdot k - i\varepsilon)(n\cdot k + i\varepsilon)}
                 + \cdots,
\end{eqnarray}
where we use the subscript $S$ to denote the contribution from the soft gluon. We have here  $q^+ =k_A^+$ and $q^-= k_B^-$. We replace the vector $n$ with $u$ and $l$ with $v$. The color trace can be taken out by noting
that the density matrices  are diagonal in color space. Therefore,
\begin{eqnarray}
W^{\mu\nu}\biggr\vert_{6aS}  &\approx &   g_s^2 \int d^2 \tilde k_{A\perp} d^2 \tilde k_{B\perp}  \frac{d^4 k}{(2\pi)^4} 2\pi \delta (k ^2)
\delta^2 (k_{B\perp}+ k_{A\perp}  -k_\perp -q_\perp )
\nonumber\\
&&  {\rm Tr} \biggr [ {\bar {\mathcal M}}(\tilde k_B) \gamma^\mu
                {\mathcal M}(\tilde k_A) \gamma^\nu  \biggr ]  \frac{1}{N_c} {\rm Tr} ( T^a T^a ) \frac{ u\cdot v} { (v\cdot k - i\varepsilon)(u\cdot k + i\varepsilon)}
                 + \cdots.
\label{FIG6C}
\end{eqnarray}
This contribution can be represented with Fig.6c.  We note here that this contribution is not contained in Fig.1c.
It can be subtracted with a soft factor as shown in \cite{CSS,JMY}.
The need of such a soft factor is also necessary for one-loop virtual correction as shown later.

\par
If we want to subtract the soft region of the gluon from the contribution from Fig.6a,
we should notice that the collinear contribution in Eq.(\ref{W6AA})  also contains the contribution from the
soft region. One can easily find that the soft contribution in Eq.(\ref{W6AA}) is  exactly the same
as the soft contribution from Fig.6a given in Eq.(\ref{FIG6C}).
We can re-write the contribution from Fig.6a as:
\begin{equation}
W^{\mu\nu}\biggr\vert_{6a} =  \biggr (  W^{\mu\nu}\biggr\vert_{6a} - W^{\mu\nu}\biggr\vert_{6aA} -W^{\mu\nu}\biggr\vert_{6aB}+ W^{\mu\nu}\biggr\vert_{6aS} \biggr )  +  \biggr  ( W^{\mu\nu}\biggr\vert_{6aA} + W^{\mu\nu}\biggr\vert_{6aB} -  W^{\mu\nu}\biggr\vert_{6aS} \biggr ) ,
\label{SUBS}
\end{equation}
where $W^{\mu\nu}\vert_{6aB} $  is the contribution from the region in which the gluon is collinear to $P_B$. This contribution contains the same soft contribution as given by Fig.6c.  In the first
$(\dots)$  of Eq.(\ref{SUBS})  there are no soft- and collinear contributions. In fact
the leading contribution from the first $(\cdots)$ is zero as discussed before.
In the second $(\cdots)$ the collinear
contributions are  already included in Fig.1c, hence give no contribution to one-loop factorization.

\par

\par
\begin{figure}[hbt]
\begin{center}
\includegraphics[width=7cm]{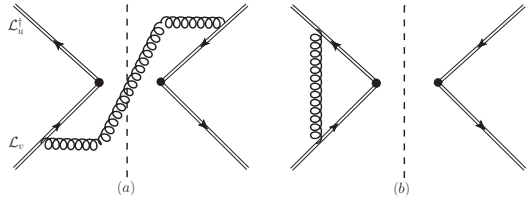}
\end{center}
\caption{Part of diagrams for the one-loop correction of the soft factor.    }
\label{dg8}
\end{figure}

\par
We define the soft factor  for the subtraction of soft gluons as:
\begin{equation}
  \tilde S(k_\perp) =  \int \frac {d^2\xi_\perp}{(2\pi)^2} e^{-i\xi_\perp \cdot k_\perp } \frac{N_c}
   { \langle 0 \vert {\rm Tr}  \left [ {\mathcal L}^\dagger_v ( \xi_\perp,-\infty) {\mathcal L}_u (\xi_\perp,-\infty)
      {\mathcal L}_u^\dagger (0,-\infty) {\mathcal L}_v (0,-\infty) \right ]  \vert 0\rangle }  .
\end{equation}
At tree-level one has:
\begin{equation}
   \tilde S^{(0)} (k_\perp) = \delta^2(k_\perp).
\end{equation}
At one-loop it receives contributions from Fig.7a and Fig.7b. The total one-loop correction is the sum of the two diagrams,
their conjugated diagrams and those diagrams for self-energy of gauge links and for one gluon exchange between ${\mathcal L}_u$
and ${\mathcal L}_v^\dagger$ and between ${\mathcal L}_v^\dagger$ and ${\mathcal L}_v$.
The contribution from Fig.7a is:
\begin{equation}
\tilde  S(\ell_\perp)\biggr\vert_{7a} = - g_s^2 \frac{1}{N_c} {\rm Tr}\left ( T^a T^a \right ) \int\frac{d^4 k}{(2\pi)^4} (2\pi) \delta (k^2) \delta^2 (k_\perp-\ell_\perp)
   \frac{ u\cdot v}{(v\cdot k - i\varepsilon)(u\cdot k + i \varepsilon)},
\label{S7A}
\end{equation}
this contribution is similar to the part factorized out in Eq.(\ref{FIG6C}).
\par
With the defined soft factor we  modify the  factorization at tree-level in Eq.(\ref{LTREE}) as
\begin{eqnarray}
W^{\mu\nu}   \approx     \int d^2 k_{A\perp} d^2 k_{B_\perp} d^2 \ell_\perp  \tilde S(\ell_\perp) {\rm Tr} \biggr  [ \gamma^\mu  {\mathcal M}(x, k_{A\perp} )   \gamma^\nu \bar{\mathcal M}(y, k_{B\perp} )  \biggr ]
  \delta^2 ( k_{A\perp}+k_{B\perp} + \ell_\perp-q_\perp ).
\label{LTREES}
\end{eqnarray}
The modification will not affect the tree-level factorization. With the modification one can see that the  soft contribution
in the second $(\cdots)$  of  Eq.(\ref{SUBS}) is included in the soft factor $\tilde S$.
Hence, Fig.6c will not contribute to the one-loop factorization. We can conclude
that with the modification Fig.6 will not contribute to perturbative coefficients in the one-loop TMD factorization.
\par
The last diagram needs to be analyzed for the real correction is the conjugated diagram of Fig.6.
The analysis is similar. The conclusion remains the same as that for the diagrams analyzed  here.  Therefore, we conclude that the real correction will not contribute to perturbative coefficients. This result
may be extended beyond one-loop level.

\par\vskip20pt
\par
\begin{figure}[hbt]
\begin{center}
\includegraphics[width=12cm]{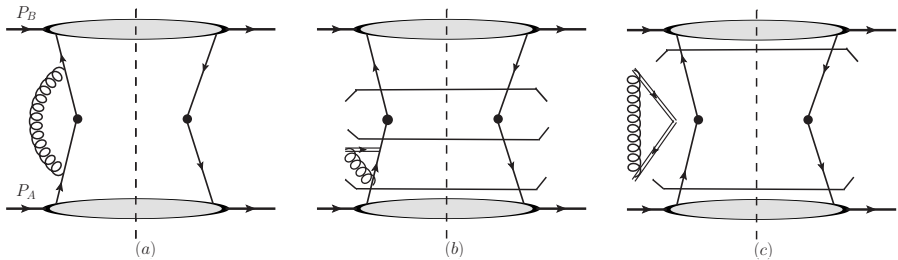}
\end{center}
\caption{The diagram for the virtual correction.    }
\label{dg8}
\end{figure}

\par\vskip20pt
\noindent
{\large \bf 5. One-Loop Virtual Correction and the Main Result }
\par\vskip5pt
In this section we  first analyze the correction from the virtual part and then we discuss the obtained result.  For the virtual diagram there are two diagrams to be studied.  One is given in Fig.8a. Another
is the conjugated one of Fig.8a.
After making the approximation for the lowest- and upper part,  one has
the contribution from Fig.8a:
\begin{eqnarray}
W^{\mu\nu}\biggr\vert_{8a} &\approx & g_s^2 \int d^3\tilde k_A d^3 \tilde k_B \frac{d^4 k}{(2\pi)^4} \delta^4 (\tilde k_A +\tilde k_B -q)
\frac{-i}{k^2+i\varepsilon}
\nonumber\\
      && \cdot {\rm Tr} \biggr [ \bar {\mathcal M } (\tilde k_B) \gamma^\rho T^a \frac{\gamma\cdot (-k-\bar k_B)}{(k+\bar k_B)^2+i\varepsilon}
           \gamma^\mu \frac{\gamma\cdot (\bar k_A -k)}{(\bar k_A -k)^2 +i\varepsilon} \gamma_\rho T^a {\mathcal M}(\tilde k_A) \gamma^\nu \biggr ],
\end{eqnarray}
where we have made the parton model approximation for the two parton lines leaving the part of $h_A$ and $h_B$. We have also  taken the limit $q_\perp \ll Q$ as explained for Fig.3a, i.e., the parton lines
from the part of $h_A$ or $h_B$ carries the momentum $\bar k_A$ or $\bar k_B$  as given in Eq.(\ref{COMEN}), respectively.
The gluon carries the momentum $k$. There are divergences
in this contribution. The contribution is divergent in three regions of $k$: The exchanged gluon is collinear to $P_A$ or to $P_B$, and the gluon is soft. One can work out the contributions from the three regions.
\par
In the case that $k$ is collinear to $P_A$ or $\bar k_A$,  with some algebra we can write the collinear contribution in the form:
\begin{eqnarray}
W^{\mu\nu}\biggr\vert_{8aA}   &\approx & \int d^2\tilde k_{A\perp} d^2 \tilde k_{B\perp} \delta^2 ( k_{A\perp} + k_{B\perp} -q_\perp)
    {\rm Tr} \biggr \{ {\bar{\mathcal M}}(\tilde k_B) \gamma^\mu
\nonumber\\
      && \cdot   \biggr [ \frac{g_s^2 C_F}{4} \int\frac{d^4 k}{(2\pi)^4} \frac{-i}{k^2+i\varepsilon} \frac{- u^\rho}{u\cdot k + i\varepsilon}   \biggr (\gamma^- \gamma^+ \biggr )
         \frac{\gamma\cdot (\bar k_A  -k)}{(\bar k_A-k)^2+i\varepsilon}  \gamma_\rho {\mathcal M}(\tilde k_A) \biggr ( \gamma^+ \gamma^- \biggr )\gamma^\nu  \biggr ] \biggr \},
\nonumber\\
\end{eqnarray}
where we have used the fact that  ${\bar{\mathcal M}}$ and ${\mathcal M}$ are diagonal in color space and replaced the vector $n$ with $u$ as before.  This contribution can be represented
by Fig.8b. The part in $ [\cdots ]$ is just the contribution  from the part of the diagram in Fig.8b between the two hooked  lines in the lower-half part.  This contribution is already included in Fig.1c. To obtain the true
correction to Fig.1c, this contribution should be subtracted.
\par
Similarly, we obtain the contribution from the region where $k$ is collinear to $P_B$:
\begin{eqnarray}
W^{\mu\nu}\biggr\vert_{8aB} &\approx & \int d^2\tilde k_{A\perp} d^2 \tilde k_{B\perp} \delta^2 ( k_{A\perp} + k_{B\perp} -q_\perp)
    {\rm Tr} \biggr \{   \biggr [\frac{g_s^2 C_F}{4} \int\frac{d^4 k}{(2\pi)^4} \frac{-i}{k^2+i\varepsilon} \frac{-v^\rho}{v\cdot k - i\varepsilon}
\nonumber\\
  &&    \biggr ( \gamma^+ \gamma^- \biggr  )   {\bar{\mathcal M}}(\tilde k_B) \gamma_\rho \frac{\gamma\cdot (-k-\bar k_B)}{(k+\bar k_B)^2+i\varepsilon}
         \biggr (\gamma^- \gamma^+ \biggr ) \biggr ] \gamma^\mu  {\mathcal M}(\tilde k_A) \gamma^\nu   \biggr \},
\nonumber\\
\end{eqnarray}
here the part in $[ \cdots ]$ is a part ${\bar {\mathcal M}}$. The above contribution should be subtracted from Fig.8a too.
The contribution from the region where $k$ is small can also be worked out. It reads:
\begin{eqnarray}
W^{\mu\nu}\biggr\vert_{8aS} &\approx & \int d^2\tilde k_{A\perp} d^2 \tilde k_{B\perp} \delta^2 ( k_{A\perp} + k_{B\perp} -q_\perp)
    {\rm Tr} \biggr [ {\bar{\mathcal M}}(\tilde k_B) \gamma^\mu{\mathcal M}(\tilde k_A) \gamma^\nu  \biggr ]
\nonumber\\
     && \biggr [  g_s^2 C_F \int\frac{d^4 k}{(2\pi)^4} \frac{-i}{k^2+i\varepsilon} \frac{ u\cdot v}
     {(u\cdot k + i\varepsilon)(v\cdot k-i\varepsilon )} \biggr ].
\label{FIG8S}
\end{eqnarray}
This contribution is divergent and it is not subtracted by contributions of TMD parton distributions.
As discussed about various contributions of Fig.6 in the last section, the collinear contributions also contain contributions from soft gluon.
These soft contributions are exactly the same as given in Eq.(\ref{FIG8S}).  The soft contributions
will be included in  the soft factor introduced in the last section in Eq.(\ref{LTREES}).
The corresponding contribution to the soft factor $\tilde S$ is from Fig.7b and  reads:
\begin{equation}
  \tilde S(k_\perp) \biggr\vert_{7b} = -  g_s^2 C_F   \delta^2 (k_\perp) \int\frac{d^4 k}{(2\pi)^4} \frac{-i}{k^2+i\varepsilon} \frac{ u\cdot v}
     {(u\cdot k + i\varepsilon)(v\cdot k-i\varepsilon )}.
\end{equation}

\par

The true contribution from Fig.8a to one-loop TMD factorization can be obtained
after the subtraction and the integration of $k$:
\begin{eqnarray}
  &&  W^{\mu\nu}\biggr\vert_{8a}-  \biggr [ W^{\mu\nu}\biggr\vert_{8aA} +  W^{\mu\nu}\biggr\vert_{8aB} -  W^{\mu\nu}\biggr\vert_{8aS} \biggr ]
\nonumber\\
 && \approx     \left \{  \frac{\alpha_s C_F}{4\pi}
  \left [ 2\pi^2 -4
    -\ln\frac{\mu^2}{Q^2} \left ( 1+ \ln\rho^2  \right )
     -\ln\rho^2 +\frac{1}{2}\left ( \ln^2\frac{Q^2}{\zeta_v^2}
      +\ln^2\frac{Q^2}{\zeta_u^2} \right ) \right ] + \cdots \right \}
\nonumber\\
&& \cdot  \int d^2 k_{A\perp} d^2 k_{B_\perp} d^2 \ell_\perp  \tilde S(\ell_\perp) {\rm Tr} \biggr  [ \gamma^\mu  {\mathcal M}(x, k_{A\perp} )   \gamma^\nu \bar{\mathcal M}(y, k_{B\perp} )  \biggr ]
  \delta^2 ( k_{A\perp}+k_{B\perp} + \ell_\perp-q_\perp ),
 \label{FACFIG8}
\end{eqnarray}
with $\rho^2 = (2 u.v)^2/(u^2 v^2)$.  The $\cdots$ in the above stands for an imaginary part. The contribution with the imaginary part 
is cancelled when we add the contribution from the diagram which is conjugated to Fig. 8. There is no contribution from 
the imaginary part in the final result of $W^{\mu\nu}$ at leading twist. 
In Eq.(\ref{FACFIG8}),   the subtracted contributions from the
 $[\cdots ]$
in the first line are either already included in Fig.1c or in the soft factor.
We have introduced gauge links along non-light-cone directions. This results in that  in TMD parton distributions there are self-energy corrections of gauge links, corrections from gluon exchanges between ${\mathcal L}_u$ and  ${\mathcal L}_u^\dagger$ and  between
${\mathcal L}_v$ and ${\mathcal L}_v^\dagger$. These corrections are all canceled by the corresponding corrections in the soft factor in Eq.(\ref{LTREES}).
\par

From the contribution of Fig.8a, one can obtain
the contribution of the conjugated diagram of Fig.8a.
Summing all contributions from one-loop correction, we obtain TMD factorization of the hadronic tensor
at one-loop level as:
\begin{eqnarray}
W^{\mu\nu} &=&  H (Q,\zeta_u^2,\zeta_v^2) \int d^2 k_{A\perp} d^2 k_{B_\perp} d^2 \ell_\perp   \delta^2 ( k_{A\perp}+k_{B\perp} + \ell_\perp-q_\perp )
\nonumber\\
   && \quad\quad\quad \cdot  \tilde S(\ell_\perp,\rho^2 ) {\rm Tr} \biggr  [ \gamma^\mu  {\mathcal M}(x, k_{A\perp},\zeta_u^2  )   \gamma^\nu \bar{\mathcal M}(y, k_{B\perp},\zeta_v^2 )  \biggr ],
\nonumber\\
 H  &=& 1 + \frac{\alpha_s C_F}{2\pi}
  \left [ 2\pi^2 -4
    -\ln\frac{\mu^2}{Q^2} \left ( 1+ \ln\rho^2  \right )
     -\ln\rho^2 +\frac{1}{2}\left ( \ln^2\frac{Q^2}{\zeta_v^2}
      +\ln^2\frac{Q^2}{ \zeta_u^2} \right ) \right ] +{\mathcal O}(\alpha_s^2).
\label{MAIN}
\end{eqnarray}
This is our main result. In the factorized form  the dependence of every quantity on directions
of used gauge links is explicitly indicated, and the dependence on the renormalisation scale $\mu$
is suppressed. The hadronic tensor does not depend on $\mu$, $\zeta_u$ and $\zeta_v$.
From the result the one-loop correction is the same for all structure functions in TMD factorization
at leading twist or leading power.  The analysis here and in the last section can be generalized
beyond one-loop order.  The perturbative coefficient is then in fact determined
by the time-like quark form factor subtracted with contributions from TMD parton distributions
and from the soft factor. 
\par

\par
In our main result of the TMD factorization we have used the unsubtracted TMD parton distributions defined in Eq.(\ref{DENM})
and the soft factor $\tilde S$. One can also use the subtracted TMD parton distributions and the soft factor as in \cite{JMY} for the factorization without any change of the perturbative coefficient.
One can also use the recently proposed definitions of TMD parton distributions in \cite{JCTMD}. In this case, the perturbative coefficient
will be changed accordingly.
\par
With the density matrices decomposed in Eq.(\ref{DECM},\ref{DECMB}) one can obtain the hadronic tensor in terms of various structure functions. The results are lengthy. We give them in the Appendix, where we label the structure functions as
$W_{AB}^{(i)}$ with $i$ as an integer. The index $A(B)$ represents the polarization of $h_A(h_B)$,
$A=U,L$ and $T$ stand for unpolarized-, longitudinally polarized- and transversely polarized $h_A$.
There are 24 structure functions. They are factorized with various TMD parton distributions. All of them are factorized with the same perturbative coefficient $H$.

\par
In the literature the one-loop correction  only exists for $W^{(0)}_{UU,LL,TT}$ in \cite{JMYP} and
for $W_{TU}^{(1)}$ in \cite{KXY}. These results agree with ours  by taking $\zeta_u^2 = \zeta_v^2 = \rho Q^2$. The correction for the first three structure functions can be extracted
by studying the partonic scattering process $q + \bar q \to \gamma^* +X$ by replacing each initial hadron
by a single quark or antiquark. But, for remaining structure functions one can not obtain useful
results by studying the partonic process. There are many reasons for it. Taking  $W_{TU}^{(1)}$
responsible for SSA as an example, if we replace the transversely polarized hadron $h_A$ with
a transversely polarized quark and the unpolarized hadron $h_B$ with an unpolarized antiquark,
one can never get nonzero result of   $W_{TU}^{(1)}$ of the partonic process $q +\bar q \to \gamma^* +X$.  The reason is the conservation of helicity in QCD.
\par
The correction for $W_{TU}^{(1)}$ in \cite{KXY} is obtained through a nontrivial way. In the first step
one  performs a collinear factorization in the impact space with twist-3 matrix elements
introduced in \cite{EFTE,QiuSt}.  Then one adds one-loop correction. There are two relations between 
Sivers function $f_{1T}^\perp (x, k_\perp)$ and twist-3 matrix elements. The first one is to relate the second $k_\perp$-moment of Sivers function 
with twist-3 matrix elements\cite{BMP}. The second one is to relate Sivers function at large $k_\perp$ 
with twist-3 matrix elements. In this case, the relation is perturbatively determined in \cite{JQVY1,JQVY2}. Using the second relation,
 one finds the one-loop correction
to  $W_{TU}^{(1)}$ in TMD factorization. It is interesting to find that the correction to  $W_{TU}^{(1)}$ obtained in this way is the same as those to $W^{(0)}_{UU,LL,TT}$.  From our results in Eq.(\ref{MAIN}), the correction is the same
for all structure functions.  At first look it may be difficult to understand this result, because the diagrams treated
in \cite{KXY} and the diagrams treated here are different. The difference also exists with some explicit calculations
in \cite{MS1,MS2}.  We will discuss this problem in the next section.

\par\vskip20pt
\noindent
{\large\bf 6. Comparing Explicit Calculations of Multi-Parton States}
\par\vskip5pt
In this section we take SSA as an example to discuss the problem mentioned in the above with the motivation
to understand our main result. The relevant structure function is $W_{TU}^{(1)}$ given in the Appendix.
As explained, if we replace the transversely polarize $h_A$ with a transversely polarized quark, one will get zero results for the Sivers function $f_{1T}^\perp$ of the quark and the structure function $W_{TU}^{(1)}$.
As discussed in detail in \cite{MS1}, one can replace $h_A$ with a state $\vert n \rangle = c_0 \vert q \rangle
  + c_1 \vert qg \rangle +\cdots$ as a superposition of a single quark state with other states containing more than one parton.  For the unpolarized $h_B$  we replace it with an antiquark $\bar q$.
Then one can study the Sivers function $f_{1T}^\perp$ of the state $\vert n\rangle$ and $W_{TU}^{(1)}$ of the partonic process $\vert n\rangle + \bar q  \to \gamma^* + X$. Since every quantities are of parton states, one can calculate them explicitly and
directly examine factorizations.

 \par
\begin{figure}[hbt]
\begin{center}
\includegraphics[width=10cm]{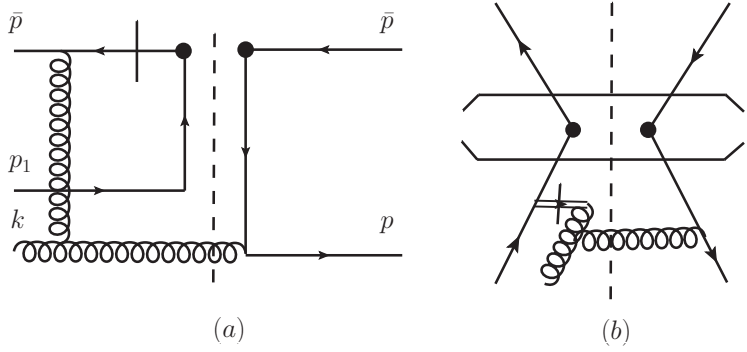}
\end{center}
\caption{(a): The diagram for SSA at leading order. (b): Factorization in the notation here.   }
\label{dg8}
\end{figure}

\par
At the leading order for the partonic process there are many diagrams\cite{MS1}.  One of them is shown in Fig.9a.  The diagrams including Fig.9a here are  already unlike the set of diagrams
given in Fig.1b in the subtractive approach.  In Fig.9a
 momenta of the partons are: $\bar p^\mu =(0,\bar p^-, 0,0)$ corresponding to the momentum of $h_B$,
$p^\mu =(p^+,0,0,0)$ corresponding to the momentum of $h_A$,  $ p_1^\mu = x_0 p^\mu$ and $k^\mu =(1-x_0) p^\mu$ are the momenta of the incoming quark and gluon, respectively.  The short bar cutting the quark propagator
means to take the absorptive part of the propagator. In fact, the short bar here is a physical cut for the absorptive part
of the left amplitude.

\par
In Fig. 9a one can nowhere to take the parton model
approximation indicated by Eq.(\ref{DENM}), or to put some hooked lines.  However, in the limit of $q_\perp/Q \ll 1$,
the approximation can be made.
Denoting the momentum and the polarization of the gluon attached to the incoming antiquark as $k_g$
and $\rho$, we have for the part involving the propagator with the bar as:
\begin{equation}
  \Gamma =   \bar v(\bar p) (-i g_s \gamma^\rho T^a ){\rm Abs} \biggr [ \frac{i\gamma\cdot (-\bar p -k_g) } {(\bar p +k_g)^2 + i\varepsilon}\biggr ]
       \gamma^\mu = \bar v(\bar p) (i g_s \gamma^\rho ) \gamma\cdot (\bar p + k_g)
       \gamma^\mu \pi\delta ((\bar p + k_g)^2),
\end{equation}
where $\gamma^\mu$ is from the electromagnetic vertex. In the limit $q_\perp /Q \ll 1$, the momentum $k_g$ is at the order $k_q^\mu \sim {\mathcal O}(\lambda^2,\lambda^2,\lambda,\lambda)$.
Hence we can approximate the above expression as:
\begin{equation}
 \Gamma \approx   \bar v(\bar p) (ig_s n^\rho T^a)  \gamma^ \mu  \pi \delta ( n \cdot k_g)
   \approx   \bar v(\bar p)   \gamma^ \mu  (i g_s u^\rho T^a )  {\rm Abs } \biggr [ \frac{i}{ u\cdot k_g +i\varepsilon} \biggr ],
\end{equation}
in the last step we have replaced the vector $n$ with the vector $u$.
Now it is clear that the contribution of Fig.9a  to $W_{TU}^{(1)}$ can be factorized as given in Fig.9b with the leading order result of $f_{1T}^\perp$ of the state $\vert n\rangle$ given in \cite{MS1} and the leading result of $\bar f_1 (y,k_\perp)$:
\begin{equation}
  f_{1T}^\perp (x,k_\perp) =
      -\frac{g_s\alpha_s}{4\pi (k_\perp^2)^2} N_c (N_c^2-1) x_0 \sqrt{2 x_0} \delta(x-x_0),  \quad\quad
      \bar f_1 (y,k_\perp) =\delta(1-y) \delta^2 (\vec k_\perp).
\label{SIQ}
\end{equation}
In the limit $q_\perp /Q \ll 1$ Fig.9a is the only diagram contributing to $W_{TU}^{(1)}$.  The above discussion indicates that
at the leading order the diagrams of multi-parton scattering reduce to the one treated in the subtractive approach
only in the limit $q_\perp /Q \ll 1$.

\par
\begin{figure}[hbt]
\begin{center}
\includegraphics[width=13cm]{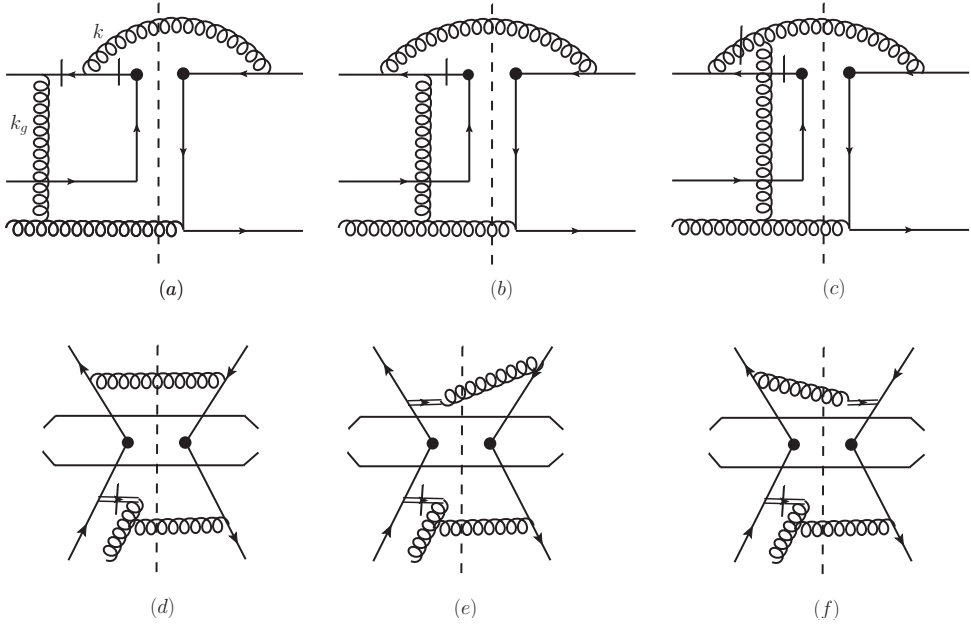}
\end{center}
\caption{(a): One diagram for the one-loop correction to Fig.9a. In this diagram the left-amplitude can have two cuts
indicated by the bars. (b,c): The same as for Fig.10a. In Fig.10b there is only one bar, in Fig.10c there are two.
(d,e,f): The factorized diagrams.  }
\label{dg8}
\end{figure}
\par
At the next-to-leading order, one needs to calculate the one-loop correction of the structure function $W_{TU}^{(1)}$,
and one-loop correction of Sivers function and the antiquark distribution to examine
the factorization of $W_{TU}^{(1)}$. The complete calculation of these corrections will be very
tedious. However,  some of the one-loop real corrections of $W_{TU}^{(1)}$ has been obtained in \cite{MS2}
for the purpose to identify the so-called soft-gluon-pole contributions in collinear factorization.  One can use these
results to provide certain understanding of the difference in diagrams  for the derivation
of our main result and for the explicit calculation presented in \cite{MS2}.
\par
In Fig.9a the antiquark line does not emit any gluon which is collinear to $\bar p$.  We consider a set of diagrams where  a gluon collinear to $\bar p$ is emitted by the antiquark lines.
The contributions from this set of diagrams are of one-loop order.
Some of such diagrams are given in Fig.10a,10b and 10c.  There are more than 10 diagrams and their conjugated diagrams.
In contrast, the corresponding diagrams analyzed in the subtractive approach in Sect.4. are only three.
The factorization of Fig.10b is rather straightforward. It is similar to Fig.5 discussed in Sect. 4.
But for other diagrams
it is difficult to see how they are factorized,  because one can not directly identify a part in those diagrams
as Sivers function. These diagrams are not similar to that given in Fig.6a.
We also note here that in Fig.10a, 10b and 10c we essentially have a Glauber gluon  represented by the vertical gluon line attached to a given diagram
everywhere where it is possible. Because it is a Glauber gluon, Ward identity may not help.
\par
In \cite{MS2} all of the diagrams mentioned in the above has been calculated in the limit $q_\perp/Q$.
The leading contribution comes from the region where the gluon emitted from an antiquark line is collinear to $\bar p$. From \cite{MS2}
we have the result of this part with $y\neq 1$ after performing the integration of a loop momentum:
\begin{equation}
W_T^{(1)}(x,y,q_\perp) \biggr \vert_y =\frac{1}{Q_\perp} W_{TU}^{(1)} \biggr \vert_y= - \frac{g_s\alpha_s^2}{ (4\pi)^2} \left (-\frac{2}{\epsilon_c} \right ) \frac{\sqrt{2 x_0}}{(q_\perp^2)^2} x_0 \delta (x-x_0) \frac{(N_c^2-1)^2}{N_c}  \biggr [  \frac{2y}{(1-y)_+} + 1-y \biggr ] +\cdots,
\label{WUTY}
\end{equation}
in \cite{MS2} only the divergent part with $d=4-\epsilon_c$ is calculated. The part with $ y=1$ is the contribution from the region where
the emitted gluon is soft. We have used the notation $\vert_y$ to denote the contribution from the collinear gluon.
This contribution should be factorized with the leading order $q_T$ given in Eq.(\ref{SIQ}) and
the one-loop real correction of the TMD parton distribution of the $\bar q$ given as:
\begin{equation}
  \bar f_1 (y,k_\perp) \biggr\vert_y = \frac{\alpha_s (N_c^2-1) }{4 \pi^2 N_c} \frac{1}{k_\perp^2} \biggr [ \frac{2y}{(1-y)_+} + (1-y) \biggr ].
\label{QBY}
\end{equation}
From the expression in TMD factorization given in Eq.(\ref{WTTF}) of the Appendix the discussed contribution should be factorized as:
\begin{equation}
W_T^{(1)}(x,y,q_\perp) \biggr\vert_y = \frac{1}{N_c} \int d^2 k_{A\perp}  d^2 k_{\perp} \frac{q_\perp \cdot  k_{A\perp}}{-Q^2_\perp}
  f_{1T}^\perp (x,k_{A\perp} ) \left ( \bar f_1 (y, k_{B\perp})\biggr\vert_y \right )
    \delta^2 (k_{A\perp} +\vec k_{B\perp} -\vec q_\perp ),
\label{CHTMD}
\end{equation}
where the soft factor is at the leading order. Using the results in Eq.(\ref{SIQ},\ref{QBY}), one should re-produce from the right-hand side
of Eq.(\ref{CHTMD}) the same divergent part given in Eq.(\ref{WUTY}).
It is easy to find it is indeed the case.  This implies that the contributions of these more than 10 diagrams
in the collinear region are factorized with the one-loop contribution of $\bar f_1 (y,k_\perp)$ with $y< 1$.
\par
This result is in fact not strange. For this set of diagrams, Fig.10b can be easily factorized as Fig.10d. The contributions from  the remaining
many diagrams in the limit of $q_\perp/Q \ll 1$ can be summed via Ward identity as the contribution given
by Fig.10e and Fig.10f, and hence can be factorized.  Therefore,  after the summation the many diagrams in the calculation
with multi-parton states in the limit $q_\perp/Q\ll 1$ are equivalent to those treated in Sect. 4. This is in accordance with the
discussion after Eq.(6), where it is clarified that $\Gamma$ and $\bar\Gamma$ only include contributions of diagrams with transverse momenta at order of $\Lambda_{QCD}$.
\par 
There is no calculation of one-loop virtual corrections for SSA with the multi-parton state, i.e., the virtual correction to Fig.9a. 
It is complicated to understand how the virtual corrections are factorized in the explicit example here. 
One complication comes from the one-loop correction of the left part of Fig.9a. In this part, the gluon exchanged between 
the gluon- and the anti-quark line is a Glauber gluon, as mentioned before. In the diagrams for one-loop correction, this Glauber 
gluon can be attached to several places, unlike in the case of the leading order, where the gluon can only be attached to one place 
as shown in Fig.9a. If one can use Ward identity for this gluon, it is rather easy to understand that the virtual corrections of the 
explicit example are factorized as studied in the last section.  But, it is not sure if Ward identity can be used here. 
Besides the corrections from diagrams by adding additionally exchanged gluon in Fig.9a, there is another class 
of diagrams which are not generated by adding one gluon line to Fig.9a, e.g., Fig.4 in \cite{MS2}. The contributions 
from these diagrams seem to be factorized with one-loop correction of Sivers function of the multi-parton state.  
To completely understand how the virtual correction in this example is factorized, a further study is needed.

\par\vskip10pt
\noindent
{\large \bf 7. Summary}
\par\vskip5pt
By using the subtractive approach we have studied TMD factorization for Drell-Yan processes beyond tree-level.
The approach is based on Feynman diagrams. In given diagrams there are nonperturbative contributions which need to be subtracted into TMD parton distributions. With the approach one can systematically construct contributions for subtracting non-perturbative effects represented by diagrams. The nonperturbative effects are only included in TMD parton distributions and the defined soft factor.  After the subtraction, one obtains the perturbative contributions. We find that at one-loop the perturbative coefficients
are the same for all 24 structure functions in TMD factorization at leading twist.
This result may be generalized beyond the one-loop level, and the perturbative coefficient is then
determined by the time-like quark form factor with the subtraction.
\par
The QCD correction can also be obtained by studying corresponding parton processes.  Replacing
each initial hadron with a single parton,  one can obtain the correction only for three structure
functions. In this case, the diagrams are similar to those treated in the subtractive approach.
For other structure functions one has to replace  each initial hadron with a multi-parton state. By studying
the scattering of multi-parton states, one can obtain the QCD corrections. However,
the diagrams of the scattering are different than those in the subtractive approach.
With existing results for the structure function responsible for SSA, one can show that
the studied diagrams in the scattering of multi-parton states in the limit $q_\perp/Q \ll 1$ are equivalent to those
in the subtractive approach after using Ward identity. 
\par
Our analysis can be straightforwardly generalized to the case of SIDIS. Hence, one expects that there is only
one perturbative coefficient for all structure functions in SIDIS, and it is determined by the space-like quark form factor
with certain subtractions. The same analysis can also be extended to TMD factorization with TMD gluon distributions
in inclusive production of Higgs,  quarkonium and two-photon system.  One may expect that
the same conclusion can be made in the case with TMD gluon distributions. Works toward this are in progress.

\vskip40pt
\renewcommand{\theequation}{A.\arabic{equation}}
\setcounter{equation}{0}


\noindent
{\large\bf Appendix: The Results of the Hadronic Tensor}
\par\vskip5pt
\par
We organize our results of the hadronic tensor as in the following: We denote the tensor as
\begin{equation}
  W^{\mu\nu} = \sum_{ (A,B)=U,L,T} W^{\mu\nu}_{AB},
\end{equation}
where the index $A=U,L$ and  $T$ denotes the contributions which do not depend on the spin of $h_A$,
the contributions depending on $\lambda_A$ and  the contributions depending on $S_{A}^\mu$,  respectively.  $B$ denotes the similar contributions related to the spin of $h_B$.
The detailed results  can be represented in terms of structure functions.
We assume that the polarization of leptons in the final state is not observed. In this case, one can only measure
the symmetric part of the hadronic tensor.
We will only give the symmetric part of the hadronic tensor.
We use the following notations:
\begin{eqnarray}
&& \int_\perp  F = \frac{1}{N_c} H \int d^2 k_{A\perp} d^2 k_{B\perp} d^2 \ell_\perp \tilde S(\ell_\perp)
  \delta^2 (k_{A\perp} +k_{B\perp} + \ell_\perp -q_\perp) F ,
\nonumber\\
 &&  \tilde q_\perp^\alpha =\epsilon_\perp^{\alpha \beta} q_{\perp\beta},
  \quad\quad Q_\perp = \sqrt{ -q_\perp\cdot q_\perp} >0 .
\end{eqnarray}
$H$ is the perturbative coefficient given in Eq.(\ref{MAIN}).
We denote the symmetric- and traceless tensor built from two transverse vectors as:
\begin{equation}
 A^{\{ \mu}_\perp B_\perp^{\nu \}} = A_\perp^\mu B_\perp^\nu + A_\perp^\nu B_\perp^\mu
    - g_\perp^{\mu\nu} A_\perp\cdot B_\perp .
\end{equation}
In the following we give the results of structure functions in different case.
\par
For  the case with $AB=UU$:
\begin{eqnarray}
W_{UU}^{\mu\nu} = -g_\perp^{\mu\nu}  W_{UU}^{(0)}  + q_\perp^{\{\mu } q_\perp^{\nu\}} \frac{1}{Q_\perp^2}  W_{UU}^{(1)},
\end{eqnarray}
with the following structure functions:
\begin{eqnarray}
 W_{UU}^{(0)} &=&  \int_\perp  f_1(x,k_{A\perp}) \bar f_1(y,k_{B\perp}),
 \nonumber\\
 W_{UU}^{(1)} &=& - \int_\perp  \tilde C_1 (k_{A\perp},k_{B\perp}, q_\perp) \frac{h_1^\perp (x,k_{A\perp} ) \bar{h}_1^\perp(y,k_{B\perp})}{M_A M_B}.
\end{eqnarray}
\par
For $AB=LU$:
\begin{eqnarray}
W^{\mu\nu}_{LU} = \lambda_A  q_\perp^{ \{\mu} \tilde q_\perp^{\nu\}}   \frac{1}{Q_\perp^2} W_{LU}^{(0)},
\end{eqnarray}
with
\begin{equation}
 W_{LU}^{(0)} =  -\int_\perp \tilde C_1 (k_{A\perp}, k_{B\perp}, q_\perp ) \frac{h_{1L}^\perp(x, k_{A\perp}) \bar{h}_{1}^\perp(y, k_{B\perp})}{M_A M_B}.
\end{equation}
\par
For $AB=TU$:
\begin{eqnarray}
W^{\mu\nu}_{TU} &=& -g_\perp^{\mu\nu}  \frac{S_{A}\cdot \tilde q_{\perp}}{Q_\perp}
   W_{TU}^{(1)}  + S^{\{ \mu} _{A}  \tilde q_\perp^{\nu \}}  \frac{1}{Q_\perp} W_{TU}^{(2)} +q_\perp^{\{ \mu }\tilde q_\perp^{\nu\}} \frac{S_{A}\cdot q_\perp }{Q_\perp^3}  W_{TU}^{(3)},
\nonumber\\
W_{TU}^{(1)} &=& \int_\perp  \left  [ \frac{ q_\perp\cdot k_{A\perp} } {-Q_\perp M_A} \right ]
      f_{1T}^\perp (x, k_{A\perp}) \bar f_1( y, k_{B\perp}),
\nonumber\\
W_{TU}^{(2)} &=& \int_\perp  \biggr \{ \left [ \frac{ k_{B\perp}\cdot q_\perp}{Q_\perp M_B}  \right ]
        h_{1T} (x, k_{A\perp}) \bar{h}_1^\perp (y, k_{B\perp})
\nonumber\\        
     && +    Q_\perp C_2 (k_{A\perp},k_{B\perp},q _\perp )  \frac{h_{1T}^\perp (x,k_{A\perp}) \bar{h}_1^\perp(y,k_{B\perp})}{M_A^2M_B} \biggr \},
\nonumber\\
W_{TU}^{(3)} &=&  \int_\perp    -2 Q_\perp C_1(k_{A\perp},k_{B\perp},q _\perp )  \frac{h_{1T}^\perp(x,k_{A\perp}) \bar{h}_1^\perp(y,k_{B\perp})}{M_A^2M_B} ,
\label{WTTF}
\end{eqnarray}
\par
For $AB=LL$:
\begin{eqnarray}
W_{LL}^{\mu\nu} &=& \lambda_A \lambda_B \left [ -g_{\perp}^{\mu\nu} W_{LL}^{(0)}  - q_\perp^{\{ \mu} q_\perp^{\nu \}} \frac{1}{Q_\perp^2}  W_{LL}^{(1)} \right ] ,
\nonumber\\
 W_{LL}^{(0)} &=&  -\int_\perp g_{1L} (x,k_{A\perp})
   \bar g_{1L}(y, k_{B\perp} ),
\nonumber\\
W_{LL}^{(1)} &=&   \int_\perp  \tilde C_1 (k_{A\perp},k_{B\perp}, q_\perp) \frac{h_{1L}^\perp (x,k_{A\perp})
  \bar{h}_{1L}^\perp  (y, k_{B\perp})}{M_A M_B},
\end{eqnarray}
\par
For $AB=TT$:
\begin{eqnarray}
W_{TT}^{\mu\nu} &=& -S_A^{\{ \mu } S_B^{\nu\}}   W_{TT}^{(0)} +
      g_\perp^{\mu\nu} \biggr [ \frac{1}{-Q_\perp^2 }  q_\perp \cdot S_A q_\perp\cdot S_B W_{TT}^{(1)}
   + S_A \cdot S_B W_{TT}^{(2)} \biggr ]
\nonumber\\
    &&  +\frac{S_A\cdot q_\perp}{Q_\perp^2} q_\perp^{\{\mu} S_B^{\nu\}}  W_{TT}^{(3)}
 + \frac{S_B\cdot q_\perp}{Q_\perp^2}  q_\perp^{\{\mu} S_A^{\nu\}} W_{TT}^{(4)}
 +  q_\perp^{\{\mu} q_\perp^{\nu\}}  \frac{S_A\cdot q_\perp S_B\cdot q_\perp}{Q_\perp^4 }  W_{TT}^{(5)},
\nonumber\\
W_{TT}^{(0)} &=&  \int_\perp  \biggr [ h_{1T} (x,k_{A\perp}) \bar{h}_{1T} (y,k_{B\perp}) + \tilde C_2 (k_{A\perp},k_{A\perp},q_\perp)
\frac{h_{1T}^\perp (x,k_{A\perp}) \bar{h}_{1T}(y,k_{B\perp})}{M_A^2}
\nonumber\\
      &&   \quad + (D_3-D_1) (k_{A\perp},k_{B\perp},q_\perp)    \frac{h_{1T}^\perp(x,k_{A\perp} )   \bar h_{1T}^\perp (y,k_{B\perp})}{M_A^2 M_B^2}
\nonumber\\     
      && \quad  + \tilde C_2 (k_{B\perp}, k_{B\perp}, q_\perp) \frac{h_{1T} (x, k_{A\perp}) \bar h_{1T}^\perp (y, k_{B\perp})}{M_B^2}\biggr ],
\nonumber\\
W_{TT}^{(1)} &=& \int_\perp  \tilde C_1 (k_{A\perp},k_{B\perp},q_\perp)  \biggr [  -\frac{g_{1T}(x,k_{A\perp} ) \bar g_{1T} (y,k_{B\perp})}{M_A M_B}
   +   \frac{f_{1T}^\perp(x,k_{A\perp} ) \bar{f}_{1T}^\perp (y,k_{B\perp})}{M_A M_B} \biggr] ,
\nonumber\\
W_{TT}^{(2)} &=& \int_\perp  \biggr [ \tilde C_2 (k_{A\perp},k_{B\perp},q_\perp)  \biggr (-\frac{f_{1T}^\perp(x,k_{A\perp} ) \bar{f}_{1T}^\perp (y,k_{B\perp})}{M_A M_B} - \frac{g_{1T} (x,k_{A\perp}) \bar{g}_{1T}(y,k_{B\perp})}{M_A M_B} \biggr )
\nonumber\\
  && \quad -  \tilde C_1 (k_{A\perp},k_{B\perp},q_\perp)   \frac{f_{1T}^\perp(x,k_{A\perp} ) \bar{f}_{1T}^\perp (y,k_{B\perp})}{M_A M_B} \biggr] ,
\nonumber\\
W_{TT}^{(3)} &=& \int_\perp \tilde C_1 (k_{A\perp},k_{A\perp},q_\perp)
 \frac{h_{1T}^\perp(x,k_{A\perp})  \bar{h}_{1T} (y,k_{B\perp})}{M_A^2} 
\nonumber\\ 
 && \quad + (3D_1+D_4) (k_{A\perp},k_{B\perp},q_\perp) \frac{h_{1T}^\perp(x,k_{A\perp} )\bar{h}_{1T}^\perp (y,k_{B\perp})}{M_A^2 M_B^2}  ,
\nonumber\\
W_{TT}^{(4)} &=& \int_\perp   \tilde C_1 (k_{B\perp},k_{B\perp},q_\perp)  \frac{h_{1T}(x,k_{A\perp} )\bar{h}_{1T}^\perp (y,k_{B\perp})}{M_B^2}
\nonumber\\  
   && \quad + (3D_1+D_5) (k_{A\perp},k_{B\perp},q_\perp)    \frac{h_{1T}^\perp(x,k_{A\perp} )   \bar{h}_{1T}^\perp (y,k_{B\perp})}{M_A^2 M_B^2} ,
\nonumber\\
W_{TT}^{(5)} &=& \int_\perp   -D_2 (k_{A\perp},k_{B\perp},q_\perp)
\frac{h_{1T}^\perp(x,k_{A\perp} )   \bar{h}_{1T}^\perp (y,k_{B\perp})}{M_A^2 M_B^2}.
\end{eqnarray}

\par
For $AB=TL$:
\begin{eqnarray}
W_{TL}^{\mu\nu} &=&  \lambda_B \left [ g_{\perp}^{\mu\nu} \frac {S_A\cdot q}{Q_\perp}  W_{TL}^{(1)} - q^{\{\mu} q^{\nu\}}    \frac{ S_A\cdot q_\perp}{Q_\perp^3 }   W_{TL}^{(2)}
   + S_A^{\{ \mu} q^{\nu\}}  \frac{1} {Q_\perp}
      W_{TL}^{(3)}
 \right ] ,
\nonumber\\
 W_{TL}^{(1)} &=&  \int_\perp\biggr (  \frac{k_{A\perp}\cdot q_\perp} {Q_\perp M_A} \biggr ) g_{1T} (x,k_{A\perp}) \bar{g}_{1L} (y, k_{B\perp})
\nonumber\\
W_{TL}^{(2)} &=&   \int_\perp  2 Q_\perp C_1(k_{A\perp},k_{B\perp},q _\perp ) \frac{ h_{1T}^\perp(x,k_{A\perp})\bar{h}_{1L}^\perp (y, k_{B\perp})}{M_A^2 M_B},
\nonumber\\
W_{TL}^{(3)} &=&   \int_\perp    \biggr [ Q_\perp C_2 (k_{A\perp},k_{B\perp},q _\perp )  \frac{h_{1T}^\perp(x,k_{A\perp})
\bar{h}_{1L}^\perp (y, k_{B\perp})}{M_A^2 M_B}
    +  \frac{k_{B\perp}\cdot q_\perp}{Q_\perp M_B} h_{1T} (x,k_{A\perp}) \bar{h}_{1L}^\perp (y, k_{B\perp}) \biggr ].
\nonumber\\
\end{eqnarray}
There are 24 structure functions in TMD factorization at leading twist. They are functions of $x,y$ and $q_\perp$.
The tensor $W_{UL}^{\mu\nu}, W_{UT}^{\mu\nu}$ can be obtained from  $W_{LU}^{\mu\nu}, W_{TU}^{\mu\nu}$
through suitable replacement.

\par
The coefficient functions $\tilde C_{1,2}$, $C_{1,2,3}$ and $D_{1,2,3,4,5}$ are given by:
\begin{eqnarray}
\tilde C_1 ( k_a,k_b,q) &=& \frac{1}{-q^2 } \biggr ( 2 q\cdot k_{a} q \cdot k_{b}
   - q^2 k_{a}\cdot k_{b } \biggr ),
\nonumber\\
\tilde C_2 ( k_{a},k_{b},q) &=& \frac{1}{ -q^2 } \biggr ( - q\cdot k_{a} q\cdot k_{b}
    + q^2 k_{a}\cdot k_{b} \biggr ),
\nonumber\\
 C_1 ( k_{a},k_{b},q)&=&  \frac{1}{2 (q^2)^2 }  \biggr [ 4 (q\cdot k_a)^2 q\cdot k_b - 2 q^2 q\cdot k_a k_a \cdot k_b
    - q^2  k_a^2 q\cdot k_b \biggr ],
\nonumber\\
C_2 ( k_{a},k_{b},q)  &=&\frac{1}{2q^2 }  k_a^2 k_b\cdot q -C_1 ( k_{a},k_{b},q),
\nonumber\\
  D_1 ( k_{a},k_{b},q)  &=& \frac{1}{(q^2)^2} \biggr [ q^2 q\cdot k_a q\cdot k_b k_a\cdot k_b-(q\cdot k_a)^2(q\cdot k_b)^2 \biggr ],
\nonumber\\
 D_2 ( k_{a},k_{b},q)  &=& \frac{1}{(q^2)^2}  \biggr [ (q^2)^2 k_a\cdot k_a k_b\cdot k_b-4q^2 q\cdot k_a q\cdot k_b k_a\cdot k_b-2q^2(q\cdot k_a)^2 k_b\cdot k_b
-2q^2(q\cdot k_b)^2 k_a\cdot k_a
\nonumber\\
&&+8(q\cdot k_a)^2(q\cdot k_b)^2 \biggr ],
\nonumber\\
 D_3 ( k_{a},k_{b}, q )  &=&\frac{1}{( q^2)^2}   \biggr [ (q^2)^2 k_a\cdot k_a k_b\cdot k_b-q^2(q\cdot k_a)^2 k_b\cdot k_b
-q^2(q\cdot k_b)^2 k_a\cdot k_a+(q\cdot k_a)^2(q\cdot k_b)^2 \biggr ],
\nonumber\\
D_4 ( k_{a},k_{b}, q )  &=&\frac{1}{( q^2)^2}   \biggr [-(q^2)^2 k_a\cdot k_a k_b\cdot k_b-q^2 q\cdot k_a q\cdot k_b k_a\cdot k_b+2q^2(q\cdot k_a)^2 k_b\cdot k_b
+q^2(q\cdot k_b)^2 k_a\cdot k_a
\nonumber\\
&&-(q\cdot k_a)^2(q\cdot k_b)^2 \biggr ],
\nonumber\\
D_5 ( k_{a},k_{b}, q )  &=&\frac{1}{( q^2)^2}   \biggr [-(q^2)^2 k_a\cdot k_a k_b\cdot k_b-q^2 q\cdot k_a q\cdot k_b k_a\cdot k_b+q^2(q\cdot k_a)^2 k_b\cdot k_b
+2q^2(q\cdot k_b)^2 k_a\cdot k_a
\nonumber\\
&&-(q\cdot k_a)^2(q\cdot k_b)^2 \biggr ],
\end{eqnarray}
with $k_a$, $k_b$ and $q$ are transverse momenta. Here, the dot product of two vectors is defined with the metric $g_\perp^{\mu\nu}$
and $q^2$= $q_\mu q_\nu g_\perp^{\mu\nu}$.  Through a carful comparison  our results by taking the tree-level result of $H=1$ agree with the existing results of factorized form factors in \cite{TaMu,AMS}.   
\par

\par\vskip40pt
\noindent
{\bf Acknowledgments}
\par
We thank Dr. A. Metz for pointing out an error about the number of the structure functions in the earlier version of our paper.
 The work of J.P. Ma is supported by National Nature
Science Foundation of P.R. China(No.11021092 and 11275244).
\par


\vskip40pt


\begin{thebibliography}{99}

\bibitem{Fac} J.C. Collins, D.E. Soper and G. Sterman, in "Perturbative Quantum Chromodynamics",
Ed.: A.H. Mueller, Singapore, World Scientific(1989), J.C. Collins, D.E. Soper and G. Sterman,
Annu. Rev. Nucl. Part. Sci. {\bf 37} (1987) 383.


\bibitem{CS} J.C. Collins and D.E. Soper, Nucl. Phys. B193 (1981) 381, Nucl. Phys. B213 (1983) 545(E),
Nucl. Phys. B197 (1982) 446, Nucl. Phys. B194 (1982) 445.

\bibitem{CSS} J.C. Collins, D.E. Soper and G. Sterman, Nucl. Phys. B250 (1985) 199.

\bibitem{JMY} X.D. Ji, J.P. Ma and F. Yuan, Phys. Rev. D71 (2005) 034005.

\bibitem{CAM} J.C. Collins and A. Metz, Phys. Rev. Lett. {\bf 93} 252001.

\bibitem{JMYP} X.D. Ji, J.P. Ma and F. Yuan, Phys. Lett. B597 (2004) 299.

\bibitem{JMYG} X.D. Ji, J.P. Ma and F. Yuan,  JHEP 0507:020,2005, hep-ph/0503015,

\bibitem{SecG1} D. Boer, W. J. den Dunnen, C. Pisano, M. Schlegel, and W. Vogelsang,  Phys. Rev. Lett. {\bf 108}, 032002
(2012), e-Print: arXiv:1109.1444 [hep-ph],
F. Dominguez, J.-W. Qiu, B.-W. Xiao, and F. Yuan, Phys. Rev. D {\bf 85}, 045003
(2012),
e-Print: arXiv:1109.6293 [hep-ph],
A. Metz and J. Zhou,  Phys. Rev. D {\bf 84},  051503
(2011) , e-Print: arXiv:1105.1991 [hep-ph].

\bibitem{BoPi} D. Boer and C. Pisano, Phys. Rev. D {\bf 86}, 094007 (2012), e-Print: arXiv:1208.3642 [hep-ph].

\bibitem{MWS} J.P. Ma, J.X. Wang and S. Zhao,  Phys. Rev. D88 (2013) 014027,
e-Print: arXiv:1211.7144 [hep-ph].

	
\bibitem{QSW} J.-W. Qiu, M. Schlegel and W. Vogelsang, Phys. Rev. Lett. {\bf 107} (2011) 062001,  
e-Print: arXiv:1103.3861 [hep-ph]. 


\bibitem{Sivers}D. Sivers, Phys. Rev. D41 (1990) 83, Phys. Rev. D43 (1991) 261.

\bibitem{MS1}  H.G. Cao, J.P. Ma and H.Z. Sang, Commun. Theor. Phys. 53 (2010) 313,
e-Print: arXiv:0901.2966 [hep-ph], J.P. Ma, H.Z. Sang  and  S.J. Zhu,  Phys. Rev. D85 (2012) 114011,
e-Print: arXiv:1111.3717 [hep-ph].
\bibitem{MS2} J.P. Ma and H.Z. Sang, JHEP 1104 (2011) 062,
e-Print: arXiv:1102.2679 [hep-ph].

\bibitem{MS22}  J.P. Ma  and  G.P. Zhang,  JHEP 1211 (2012) 156,
e-Print: arXiv:1203.6415 [hep-ph],  	

\bibitem{TW3EVO} J.P. Ma and  Q. Wang,  Phys. Lett. B715 (2012) 157,  e-Print: arXiv:1205.0611 [hep-ph].

\bibitem{JC1}  J.C. Collins  and  T.C. Rogers,  Phys. Rev. D78 (2008) 054012,
e-Print: arXiv:0805.1752 [hep-ph].

\bibitem{JC2}  J.C. Collins, T.C. Rogers  and  A.M. Stasto,  Phys. Rev. D77 (2008) 085009,
e-Print: arXiv:0708.2833 [hep-ph].

\bibitem{JC3}  J. Collins,   Phys. Rev. D65 (2002) 094016,
e-Print: hep-ph/0110113.

\bibitem{TaMu} R.D. Tangerman and P.J. Mulders,  Phys.Rev. D51 (1995) 3357,
e-Print: hep-ph/9403227.

\bibitem{AMS} S. Arnold, A. Metz and M. Schlegel, Phys.Rev. D79 (2009) 034005,
e-Print: arXiv:0809.2262.

\bibitem{TMDMT} P.J. Mulders and R.D. Tangerman, Nucl. Phys. B461 (1996) 197, e-Print: hep-ph/9510301,
Nucl. Phys. B484 (1997) 538(E).

\bibitem{TMDBM} D. Boer and P.J. Mulders, Phys. Rev. D57 (1998) 5780, e-Print: hep-ph/9711485.

\bibitem{TMDGMS} K. Goeke, A. Metz and M. Schlegel, Phys. Lett. B618 (2005) 90, e-print:he-ph/0504130.

\bibitem{BDGM} A. Bacchetta {\it et al.}, JHEP 0702 (2007) 093,
e-Print: hep-ph/0611265.

	
\bibitem{AJMY} A. Idilbi, X.-D Ji, J.-P. Ma  and F. Yuan,  Phys.Rev. D70 (2004) 074021,
e-Print: hep-ph/0406302.

\bibitem{TMDJi} X.D. Ji and F. Yuan, Phys. Lett. B543 (2002) 66, e-Print: hep-ph/0206057, 
A.V. Belitsky, X.D. Ji and F. Yuan, Nucl. Phys. B656 (2003) 165, e-Print: hep-ph/0208038. 

\bibitem{JCTMD} J. Collins,  in "\emph{Foundations of perturbative QCD}", (Cambridge University Press, Cambridge, 2011); Int. J. Mod. Phys. Conf. Ser. {\bf 04}, 85
(2011), e-Print: arXiv:1107.4123 [hep-ph].

	
\bibitem{KXY}  Z.-B. Kang, B-W. Xiao  and  F. Yuan,  Phys. Rev. Lett.  {\bf 107}  (2011) 152002,
e-Print: arXiv:1106.0266 [hep-ph], 	
W. Vogelsang and F. Yuan,  Phys. Rev. D79 (2009) 094010,
e-Print: arXiv:0904.0410 [hep-ph].

\bibitem{EFTE} A.V.~Efremov and O.V. Teryaev, Sov. J. Nucl. Phys. {\bf 36} (1982) 1,
Phys. Lett. B150 (1985) 383.

\bibitem{QiuSt} J. W. Qiu and G. Sterman, Phys. Rev. Lett {\bf 67} (1991) 2264,
Nucl. Phys. B378 (1992) 52, Phys. Rev. D59 (1998) 014004.

\bibitem{BMP} D. Boer, P.J. Mulders and F. Pijlman, Nucl. Phys. B667 (2003) 201, e-Print: hep-ph/0303034. 

\bibitem{JQVY1} X.D. Ji, J.W. Qiu, W. Vogelsang and F. Yuan, Phys. Rev. Lett. {\bf 97} (2006) 082002,
e-Print: hep-ph/0602239, Phys. Rev. D73 (2006) 094017,
e-Print: hep-ph/0604023.

\bibitem{JQVY2} X.D. Ji, J.W. Qiu, W. Vogelsang and F. Yuan,  Phys. Lett. B638 (2006) 178,
e-Print: hep-ph/0604128.



\end{thebibliography}
\end{document}